\documentclass[aps,nofootinbib,floatfix,amsmath,amssymb,groupedaddress,groupedaddress,prd,preprint]{revtex4-1}

\usepackage{hyperref}

\usepackage{graphicx}
\usepackage{siunitx}

\makeatletter
\@ifclassloaded{revtex4-1}{}{\usepackage{booktabs}}
\makeatother

\usepackage[font=small,labelfont=bf]{caption}
\usepackage{subfig}

\usepackage[english]{babel}

\usepackage{xspace}

\usepackage{paralist}

\usepackage{microtype}
\usepackage[T1]{fontenc}
\usepackage{bm} 

\usepackage{enumitem}

\newcommand{\acknowledge}{%
\begin{acknowledgements}
This work was supported in part by grants IUT23-6, CERN+,  and by the European Union
through the European Regional Development Fund and by ERDF project 3.2.0304.11-0313
Estonian Scientific Computing Infrastructure (ETAIS).
\end{acknowledgements}
\let\cite\oldcite{}
}

\newcommand{\mxnewcommand}[2]{\newcommand{#1}{\ensuremath{#2}\xspace}}

\newcommand{\xnewcommand}[2]{\newcommand{#1}{#2\xspace}}

\mxnewcommand{\gev}{\,\text{GeV}}
\mxnewcommand{\tev}{\,\text{TeV}}
\mxnewcommand{\invfb}{/\text{fb}}

\newcommand{\roots}[1]{\ensuremath{\sqrt{s}={#1}\tev}}

\mxnewcommand{\mhalf}{m_{1/2}}
\mxnewcommand{\mzero}{m_0}
\mxnewcommand{\azero}{A_0}
\mxnewcommand{\tanb}{\tan\beta}
\mxnewcommand{\sgnmu}{\text{sign}\,\mu}
\mxnewcommand{\sgnb}{\text{sign}\,b}
\mxnewcommand{\mhu}{m_{H_u}}
\mxnewcommand{\mhd}{m_{H_d}}
\mxnewcommand{\tanbsq}{\tan^2\beta}

\mxnewcommand{\msinglet}{m_S}
\mxnewcommand{\mueff}{\mu_\text{eff}}
\mxnewcommand{\sgnmueff}{\text{sign}\,\mueff}

\mxnewcommand{\mz}{M_Z}
\mxnewcommand{\msbar}{\overline{\text{MS}}}
\mxnewcommand{\invalpha}{1/\alpha_{\text{em}}(M_Z)^{\msbar}}
\mxnewcommand{\alphas}{\alpha_s(M_Z)^{\msbar}}
\mxnewcommand{\mt}{m_t^\text{Pole}}
\mxnewcommand{\mb}{m_b(m_b)^{\msbar}}

\mxnewcommand{\mts}{m_t}
\mxnewcommand{\mbs}{m_b}
\mxnewcommand{\invalphas}{1/\alpha_{\text{EM}}}
\mxnewcommand{\alphass}{\alpha_s}

\mxnewcommand{\mplanck}{M_\text{P}}
\mxnewcommand{\mgut}{M_\text{GUT}}
\mxnewcommand{\msusy}{M_\text{SUSY}}
\mxnewcommand{\mweak}{\mz}

\mxnewcommand{\cp}{\mathcal{CP}}

\newcommand{\lag}[1][]{\ensuremath{%
\mathcal{L}_\text{#1}%
}}

\newcommand{\superp}[1][]{\ensuremath{%
\mathcal{W}^\text{#1}%
}}
\mxnewcommand{\cc}{\text{c.c.}}

\renewcommand{\(}{\left(}
\renewcommand{\)}{\right)}

\mxnewcommand{\like}{\mathcal{L}}
\mxnewcommand{\prior}{\pi}
\mxnewcommand{\params}{m}
\mxnewcommand{\ev}{\mathcal{Z}}
\mxnewcommand{\point}{\vec{x}}
\mxnewcommand{\model}{\text{model}}
\mxnewcommand{\data}{\text{data}}
\mxnewcommand{\pd}{\,\prod\text{d}}

\newcommand{\priorf}[1]{\ensuremath{\prior(#1)}}
\newcommand{\likef}[1]{\ensuremath{\like(#1)}}

\newcommand{\pg}[2]{\ensuremath{p(#1\,\bm{|}\,#2)}}
\newcommand{\p}[1]{\ensuremath{p(#1)}}

\newcommand{\dpartial}[2]{\ensuremath{\frac{\partial #1}{\partial #2}}}
\newcommand{\indpartial}[2]{\ensuremath{{\partial #1}/{\partial #2}}}

\newcommand{\s}[1]{\ensuremath{\tilde{#1}}}
\newcommand{\sbar}[1]{\ensuremath{\tilde{\bar{#1}}}}
\newcommand{\sd}[1]{\ensuremath{\tilde{#1}^\dagger}}
\newcommand{\sbd}[1]{\ensuremath{\tilde{\bar{#1}}^\dagger}}
\newcommand{\neut}[1]{\ensuremath{{\chi}^0_{#1}}}

\newcommand{\mneut}[1]{\ensuremath{m_{\neut{#1}}}}

\mxnewcommand{\sigsip}{\sigma^{\text{SI}}_p}
\mxnewcommand{\abund}{\Omega h^2}
\xnewcommand{\stauc}{stau-coannihilation}
\xnewcommand{\Stauc}{Stau-coannihilation}

\newcommand{\br}[1]{\ensuremath{\text{BR}}(#1)}
\mxnewcommand{\bsg}{\br{B_s\to X_s\gamma}}
\mxnewcommand{\bsmm}{\br{B_s\to \mu\mu}}
\mxnewcommand{\btn}{\br{B_u\to \tau\nu}}

\mxnewcommand{\damu}{\delta a_\mu}
\mxnewcommand{\mw}{M_W}
\mxnewcommand{\dmbs}{\Delta M_{B_s}}
\mxnewcommand{\sineff}{\sin^{2} \theta_{\ell,\text{eff}}}

\mxnewcommand{\mh}{m_h}
\mxnewcommand{\ma}{m_A}

\mxnewcommand{\pmm}{(\mzero,\,\mhalf)}
\mxnewcommand{\pat}{(\azero,\,\tanb)}
\mxnewcommand{\pcs}{(\mneut{1},\,\sigsip)}
\mxnewcommand{\pkm}{(\kappa,\,\msinglet)}
\mxnewcommand{\pmmu}{(\mzero,\,\mu)}

\let\oldcite\cite
\renewcommand{\cite}{~\oldcite}
\newcommand{\reftable}[1]{Table~\ref{#1}} 
\newcommand{\reffig}[1]{Fig.~\ref{#1}}
 
\newcommand{\refeq}[1]{Eq.~(\ref{#1})}
\newcommand{\refsec}[1]{Sec.~\ref{#1}}
\newcommand{\refcite}[1]{Ref.\cite{#1}}
\newcommand{\see}[1]{(see \eg\cite{#1})}

\xnewcommand{\eg}{\textit{e.g.,}}
\xnewcommand{\ie}{\textit{i.e.,}}
\xnewcommand{\latincf}{\textit{c.f.,}}
\newcommand{\latin}[1]{\textit{#1}}
\mxnewcommand{\dash}{\text{, }}
\newcommand{\ic}[1]{``#1''}

\newcommand{\beq}{\begin{equation}}
\newcommand{\eeq}{ \end{equation}}

\begin{document}

\title{Is the CNMSSM more credible than the CMSSM?}

\author{Andrew Fowlie}
\email{Andrew.Fowlie@KBFI.ee}
\affiliation{National Institute of Chemical Physics and Biophysics, Ravala 10,
Tallinn 10143, Estonia}

\date{\today}

\begin{abstract}
With Bayesian statistics, we investigate the full parameter space of the constrained \ic{next-to-minimal} supersymmetric
Standard Model (CNMSSM) with naturalness priors, which were derived in a previous work.
In the past, most Bayesian analyses of the CNMSSM ignored naturalness of the electroweak (EW) scale by making
prejudicial assumptions for parameters defined at the EW scale. 
We test the CNMSSM against the CMSSM with Bayesian evidence, which, with naturalness priors, incorporates a penalty for fine-tuning of the EW scale. 
With the evidence, we measure credibility with respect to all measurements, including the EW scale and LHC direct searches. 
We find that the evidence in favor of the CNMSSM versus the CMSSM is \ic{positive} to \ic{strong} but that if one ignores the $\mu$-problem, the evidence is \ic{barely worth mentioning} to \ic{positive.} 
The $\mu$-problem significantly influences our findings.
Unless one considers the $\mu$-problem, the evidence in favor of the CNMSSM versus the CMSSM is at best \ic{positive,} which is two grades below \ic{very strong.}
We, furthermore, identify the most probable regions of the CMSSM and CNMSSM parameter spaces and examine prospects for future discovery at hadron colliders.
\end{abstract}

\maketitle

\section{\label{Section:Introduction}Introduction} 
The Standard Model (SM) contains a well-known \ic{hierarchy problem}\cite{Gildener:1976ai,Susskind:1978ms}. The problem has two puzzling facets:%
\begin{inparaenum}[(1)]
\item\label{hp:magnitude} why is the magnitude of the electroweak (EW) scale much less than the Planck scale, $\mweak\ll\mplanck$?~and 
\item\label{hp:stability} why is the EW scale stable despite massive quadratic corrections, $\Delta\mweak^2\sim\mplanck^2$? 
\end{inparaenum}

Weak-scale supersymmetry (SUSY)\cite{Salam:1974yz,Haber:1984rc,Nilles:1983ge} solves the \ic{stability} aspect of the hierarchy problem by positing a \ic{mirror} of the SM fields with spins differing by one-half. Massive quadratic corrections from scalars cancel with identical corrections from fermions \see{Martin:1997ns,Baer:2006rs,Dine:2007zp}. Because residual corrections are similar to the SUSY breaking scale, $\Delta\mweak^2\sim\msusy^2$, the SUSY breaking scale should be close to the EW scale\cite{Barbieri:1987fn,Ellis:1986yg}.

Minimal SUSY, however, aggravates the \ic{magnitude} aspect of the hierarchy problem. \ic{Supersymmetrizing} the SM with minimal field content, the EW scale is function of a SUSY breaking scale, \mhu, and a SUSY preserving scale, $\mu$, 
\beq
\label{Eqn:MZ}
\frac12 \mz^2 \simeq -\mu^2 - \mhu^2|_{\text{EW}}
\eeq
where $\mu$ is protected from massive quadratic corrections by a supersymmetric non-renormalization theorem but is unrelated to a symmetry breaking scale, whereas the SUSY breaking up-type Higgs mass, $\mhu^2$, receives massive radiative corrections proportional to the supersymmetric top (stop) mass, $\Delta\mhu^2\sim m_{\s{t}}^2$.\footnote{In \refeq{Eqn:MZ}, $\mhu^2|_{\text{EW}}$ is negative. The quantity $\mhu^2$ is a parameter in the soft-breaking Lagrangian; it is not the square of a parameter.}  

This is the \ic{$\mu$-problem}\cite{Kim:1983dt}. It would be preferable if the EW scale were a function of only the SUSY breaking scale so that explaining the magnitude of the EW scale would be equivalent to explaining the magnitude of the SUSY breaking scale, which presumably originates from a hidden sector. This is realized with an extra gauge singlet superfield\cite{Fayet:1974pd}; the $\mu$-parameter is generated spontaneously by SUSY breaking parameters. 

This picture is, however, spoiled by experimental results from the Large Hadron Collider (LHC) that suggest that the SUSY breaking scale is not close to the EW scale, including the measurement of the Higgs mass $\mh\simeq126\gev$\cite{Beringer:1900zz,Chatrchyan:2012ufa,Aad:2012tfa} and 
the absence of SUSY in ATLAS\cite{TheATLAScollaboration:2013fha} and CMS\cite{Chatrchyan:2014lfa} searches.
In minimal supersymmetric models \see{Martin:1997ns},
\beq
\mh^2 \simeq \cos^2 2\beta\,\mz^2 + \Delta\mh^2,
\eeq
where $\tanb = \langle H_u \rangle/\langle H_d \rangle$ and the loop-corrections
\beq
\Delta\mh^2=\frac{3}{4\pi^2} \cos^2\alpha\, y_t^2 m_t^2 \ln\left(\frac{m_{\tilde{t}}^2}{m_t^2}\right).
\eeq
Because $\mh\simeq126\gev$, the loop-corrections, $\Delta\mh$, and thus stop masses, $m_{\tilde{t}}$, must be appreciable. Heavy stops \ic{poison} the prediction for the EW scale in \refeq{Eqn:MZ}. By contributing radiatively to $\mhu^2$, heavy stops result in $-\mhu^2 \gg \mz^2$.

The separation between the SUSY breaking scale and the EW scale in \ic{next-to-minimal} models, however, could be smaller than that in minimal models. In \ic{next-to-minimal} models \see{Maniatis:2009re,Ellwanger:2009dp}, there is an additional tree-level contribution to the Higgs mass;
\beq
\label{Eqn:N_Higgs}
\mh^2 \simeq \cos^2 2\beta\,\mz^2 + \lambda^2 v^2 \sin^2 2\beta+ \Delta\mh^2,
\eeq
where $\lambda$ originates from a cubic interaction in the superpotential and $v\simeq174\gev$. The loop-corrections and thus stop masses in \ic{next-to-minimal} models could be smaller than those in minimal models, because of the extra tree-level contribution to the Higgs mass\cite{BasteroGil:2000bw,Ellwanger:2011mu,Perelstein:2012qg,King:2012tr,Kim:2013uxa,Gherghetta:2012gb,Agashe:2012zq}.

Let us examine both facets of the hierarchy problem, including the $\mu$-problem, in a \ic{next-to-minimal} and in a minimal SUSY model in light of LHC results. With Bayesian statistics, we will calculate whether a \ic{next-to-minimal} model is more credible than a minimal model, and if so, we will quantify its superiority with Bayesian evidence. With the Bayesian posterior density, we will find the most probable regions of their parameter spaces in light of experimental data and Bayesian naturalness considerations. We will show that the $\mu$-problem significantly influences our findings.

\section{\label{Section:Models}Models}
We consider two models, the CMSSM and the CNMSSM, defined below to clarify our parameterization. Our notation is similar to that of \refcite{Martin:1997ns}.

\subsection{CMSSM}
Our minimal model is the Constrained Minimal Supersymmetric SM (CMSSM)\cite{Chamseddine:1982jx,Arnowitt:1992aq,Kane:1993td}. The model's superpotential is
\beq
\label{Eqn:MSSM}
\superp[MSSM] = \bar u \bm{y_u} Q H_u - \bar d \bm{y_d} Q H_d - \bar e \bm{y_e} L H_d + \mu H_u H_d.
\eeq
The model's soft-breaking Lagrangian at the Grand Unification (GUT) scale is
\begin{align}
\begin{split}
 \label{Eqn:MSSM_soft}
\lag[soft]^\text{MSSM} = &-\frac12 \mhalf \(\s{b}\s{b}+\s{W}\s{W}+\s{g}\s{g} + \cc\)\\
&-\mzero^2 \(\sd{Q}\s{Q}+\sd{L}\s{L}+\sbar{u}\sbd{u}+\sbar{d}\sbd{d}+\sbar{e}\sbd{e}+H_u^*H_u+H_d^*H_d\)\\
&-\azero \(\sbar{u}\bm{y_u}\s{Q}H_u-\sbar{d}\bm{y_d}\s{Q}H_d-\sbar{e}\bm{y_e}\s{Q}H_d + \cc\)\\
&-b H_u H_d + \cc
\end{split}
\end{align}
Thus the model is described by five parameters: four SUSY breaking parameters,
\beq
\text{\mhalf, \mzero, \azero, and $b$},
\eeq
and the $\mu$-parameter in the superpotential.\footnote{The CMSSM is also described by Yukawa couplings in the superpotential. We consider the Yukawa couplings in the CMSSM and CNMSSM as \ic{nuisance} parameters: model parameters that are not of particular interest.} In a phenomenological parameterization of the CMSSM, $b$ and $\mu^2$ are traded for $\tan\beta$ and $\mz$ via EW symmetry breaking conditions.

The CMSSM contains two Higgs doublets with eight real degrees of freedom. In EW symmetry breaking, the $W$- and $Z$-bosons \ic{eat} three degrees of freedom from the Higgs doublets. The five remaining degrees of freedom are equivalent to five physical Higgs bosons: a light SM-like Higgs, $h$, a heavy neutral Higgs, $H$, a heavy charged Higgs, $H^\pm$, and a neutral \cp-odd Higgs, $A$.  

After EW symmetry breaking, off-diagonal masses \ic{mix} bino, wino and Higgsino fields into mass eigenstates called \ic{neutralinos,} \neut. The phenomenology of the four neutralinos is rich. If the lightest neutralino is the lightest supersymmetric particle (LSP) and if it cannot decay to SM particles, dark matter (DM) could be the lightest neutralino \see{2010pdmo.book..142E}.

\subsection{CNMSSM}
Our \ic{next-to-minimal} model is the Constrained \ic{Next-to-minimal} Supersymmetric SM (CNMSSM or C(M$+1$)SSM) with an extra gauge singlet superfield, $S$ \see{Maniatis:2009re,Ellwanger:2009dp}. The superpotential contains extra terms with the singlet superfield;
\beq
\label{Eqn:MSSM_super}
\superp[NMSSM] = \superp[MSSM]|_{\mu=0} + \lambda S H_u H_d + \frac13 \kappa S^3.
\eeq
The $\mu$-term that was permitted in the MSSM, a singlet bilinear and a singlet tadpole are forbidden by a discrete $\mathbb{Z}_3$ symmetry or classical scale invariance. Because the superpotential is protected by a non-renormalization theorem, such terms cannot be generated by radiative corrections. 

The model's soft-breaking Lagrangian at the GUT scale is
\begin{align}
\begin{split}
\label{Eqn:NMSSM_soft}
\lag[soft]^\text{NMSSM} = &\lag[soft]^\text{MSSM}|_{b=0}\\
&-\msinglet^2 S^*S\\
&-\azero\(\lambda SH_u H_d - \frac13\kappa S^3 + \cc\).
\end{split}
\end{align}
Bilinear and tadpole, $tS$, terms are forbidden by a discrete $\mathbb{Z}_3$ symmetry. A tadpole term would be problematical\cite{Ellwanger:1983mg}; if $S$ were a singlet under all symmetries, radiative corrections from any heavy fields would result in $t\gg\mweak^3$. Each trilinear coupling in the soft-breaking Lagrangian is proportional to the corresponding trilinear coupling in the superpotential in analogy with the MSSM in which \eg $\bm{a_u}=A_u \bm{y_u}$.

If the scalar field $S$ obtains a non-zero vacuum expectation value (VEV), the discrete $\mathbb{Z}_3$ symmetry and classical scale invariance are spontaneously broken, but SUSY is preserved as the vacuum expectation of the scalar potential remains zero, and an effective $\mu$-term $\mueff=\lambda\langle S\rangle H_u H_d$ is spontaneously generated in \refeq{Eqn:MSSM_super}. The magnitude of $\langle S \rangle$ is determined from a symmetry breaking constraint, $\indpartial{V}{S}=0$.

That the discrete $\mathbb{Z}_3$ symmetry is spontaneously broken is problematic. During spontaneous symmetry breaking, topologically stable field configurations, known as domain walls, would form at the spatial boundaries of degenerate vacua. The spatial variation in the field between the degenerate vacua represents a considerable  energy density. Because domain walls could dominate the energy density of the Universe, domain walls could spoil successful predictions of inflation and nucleosynthesis.

The NMSSM's additional gauge singlet superfield modifies the neutralino and Higgs sectors of the MSSM with two on-shell fermionic and two on-shell scalar degrees of freedom. The two scalar degrees of freedom result in two extra Higgs bosons and alter the mixing angles between the physical Higgs bosons and the gauge eigenstates. In general, the NMSSM Higgs-sector violates \cp-symmetry at tree-level. If, however, complex phases are forbidden, the Higgs sector respects \cp. There are three \cp-even neutral Higgs bosons, $H_1$, $H_2$ and $H_3$; two \cp-odd neutral Higgs bosons, $A_1$ and $A_2$; and one charged Higgs boson, $H^\pm$. Unlike in the CMSSM, in the CNMSSM several Higgs bosons could be near the EW scale. The observed Higgs boson need not be the lightest \cp-even neutral Higgs boson in the CNMSSM. Were the lightest Higgs boson's couplings small, it could have evaded searches for Higgs bosons at LEP, the Tevatron and the LHC \see{Higgs:PDG}. 

The singlet superfield's two fermionic degrees of freedom are a Majorana \ic{singlino.} After EW symmetry breaking, off-diagonal masses mix the singlino with the two neutral Higgsinos (which are mixed with the two neutral gauginos), resulting in five neutralinos. The Higgsino-singlino mixing is proportional to $\lambda$. If $\lambda$ is small, the singlino decouples, resulting in four MSSM-like neutralinos and a singlino. If the singlino soft-breaking mass, \msinglet, is substantial and the singlino is decoupled, it might be difficult to distinguish the MSSM and NMSSM neutralino sectors.

The model is described by six parameters: four SUSY breaking parameters,
\beq
\text{\mhalf, \mzero, \msinglet and \azero},
\eeq
and the $\lambda$ and $\kappa$ SUSY preserving parameters in the superpotential. The number of free parameters in the CNMSSM is \emph{one greater} than that of the CMSSM. 

The singlet SUSY breaking mass is not unified at the GUT scale, $m_S\neq\mzero$. This choice is partly pragmatic --- evolving $m_S=\mzero$ to the EW scale with correct EW symmetry breaking is difficult --- and partly theoretical\cite{Hugonie:2007vd,Djouadi:2008uj}. Suppose that at the Planck scale, $\mplanck\sim10^{18}\gev$, supersymmetry breaking is mediated by gravitational interactions with a hidden sector, and that the superfields are embedded into representations of a GUT group, broken at $\mgut\sim10^{16}\gev$. If SUSY breaking is universal (as in minimal supergravity), $\msinglet=\mzero$ at the Planck scale. If the singlet superfield resides in a different representation of the GUT group, renormalization group running between the Planck scale and the GUT scale will result in non-universal SUSY breaking masses at the GUT scale\cite{Polonsky:1994sr}. Moreover, SUSY breaking interactions might discriminate the singlet from the other fields. 

\section{\label{Section:Bayes}Bayesian naturalness}
Our goal is to measure the \ic{Bayesian naturalness} of the EW scale and experimental data in the CMSSM and the CNMSSM, \eg is $\mweak\sim100\gev$ a generic prediction or does it require that the model parameters are \ic{fine-tuned?}\footnote{Where there is an important distinction between traditional and Bayesian interpretations of naturalness, in a Bayesian context, we refer to \emph{Bayesian} naturalness and credibility; whereas, in a general context, we refer to naturalness.}
To measure Bayesian naturalness, we will utilize Bayesian statistics. \refcite{Cabrera:2008tj,Cabrera:2009dm,Fichet:2012sn,Fowlie:2014xha} argued that naturalness and fine-tuning arguments are Bayesian in nature. Let us briefly recapitulate this argument \see{Fowlie:2014xha}.

In Bayesian statistics, probability is a numerical measure of our degree of belief in a proposition, rather than the frequency at which outcomes occur in repeated trials. We must calculate the probability that our model is correct, given experimental data, \eg the measured EW scale. By Bayes' theorem, we may write this probability as a function of the \textit{Bayesian evidence,} $\ev\equiv\pg{\data}{\model}$; our belief in the model prior to seeing the experimental data, $\p{\model}$; and an unknown normalization constant, $\p{\data}$;
\beq
\label{Eqn:Bayes}
\pg{\model}{\data} = \frac{\pg{\data}{\model} \times \p{\model}}{\p{\data}}.
\eeq

To eliminate the normalization constant in \refeq{Eqn:Bayes},\footnote{
Alternatively, we could assume that there exists a finite set of alternative models one of which is true, in which case we could calculate the normalization constant;
\beq
\p{\data} = \sum_i \pg{\data}{\model_i} \times \p{\model_i}.
\eeq
}
we consider the ratio of probabilities for our two models, the CMSSM and the CNMSSM;
\beq
\label{Eqn:Odds}
\underbrace{\frac{\pg{\text{CMSSM}}{\data} }{ \pg{\text{CNMSSM}}{\data}}}_{\text{Posterior odds, }\theta^\prime} = \underbrace{\frac{\pg{\data}{\text{CMSSM}} }{ \pg{\data}{\text{CNMSSM}}}}_{\text{Bayes-factor, }B} \times \underbrace{\frac{ \p{\text{CMSSM}} }{\p{\text{CNMSSM}}}}_{\text{Prior odds, }\theta}.\\
\eeq
Our \textit{prior odds}, $\theta$, is a numerical measure of our relative belief in the CMSSM over the CNMSSM, \textit{prior} to seeing the experimental data. 
The Bayes-factor, $B$, updates our \textit{prior odds,} $\theta$, with the experimental data, resulting in our \textit{posterior odds,} $\theta^\prime$. Our \textit{posterior odds}  is a numerical measure of our relative belief in the CMSSM over the CNMSSM, \textit{after} seeing the experimental data. The Bayes-factor is the ratio of the models' evidences. 

If the Bayes-factor is greater than (less than) one, the model in the numerator (denominator) is favored. The interpretation of Bayes-factors is somewhat subjective, though we have chosen the Jeffreys' scale, \reftable{Table:Jeffreys_Scale}, to ascribe qualitative meanings to Bayes-factors.  If a Bayes-factor is sufficiently large, all investigators will conclude that a particular model is favorable, regardless of their prior odds for the models. 

\begin{table}[ht]
\centering
\begin{ruledtabular}
\begin{tabular}{llll}
Grade   & Bayes-factor, $B$             & Preference for model in numerator\\
\hline
0       & $B\le1$                       & Negative\\
1       & $1<B\le3$                     & Barely worth mentioning\\
2       & $3<B\le20$                    & Positive\\
3       & $20<B\le150$                  & Strong\\
4       & $B>150$                       & Very strong\\
\end{tabular}
\end{ruledtabular}
\caption{The Jeffreys' scale for interpreting Bayes-factors\cite{nla.cat-vn759870,Kass1995}, which are ratios of evidences. We assume that the favored model is in the numerator, though this could be readily inverted.}
\label{Table:Jeffreys_Scale}
\end{table}

The Bayes-factor quantitatively incorporates a Bayesian interpretation of \ic{naturalness}\cite{2009arXiv0903.4055G,Barbieri:1987fn,Ellis:1986yg}. Consider the evidence $\ev = \pg{\data}{\model}$ a function of the data normalized to unity\cite{MacKay91}. Natural models \ic{spend} their probability mass near the obtained  data, \ie a large fraction of their parameter space agrees with the data. Complicated models squander their probability mass away from the obtained data. For example, the SM is unnatural because its generic prediction for the EW scale is $\mweak\sim\mplanck$. See \eg \refcite{Fowlie:2014xha} for elaboration.

With Bayes' theorem, it can be readily  shown that the evidence is an integral over the \emph{likelihood} --- the probability of obtaining data given a particular point in a model's parameter space, $\likef{\point}\equiv\pg{\data}{\point,\model}$ --- times the \emph{prior} --- our prior belief in the model's parameter space, $\priorf{\point}\equiv\pg{\point}{\model}$;
\beq
\label{Eqn:Evidence}
\ev = \int \likef{\point}\times\priorf{\point}\pd x.
\eeq

\subsection{Bayesian posterior}
A probability density function (PDF) for the model's parameter space in light of the experimental data --- the \emph{posterior} --- is a by-product of the calculation of the Bayesian evidence. By Bayes' theorem, the posterior density for a point \point in a model's parameter space is
\begin{align}
\label{Eq:Posterior}
\begin{split} 
\pg{\point}{\model,\,\data} &= \frac{\pg{\data}{\point,\,\model} \times \pg{\point}{\model}}{\pg{\data}{\model}}\\
&\equiv \frac{\likef{\point}\times\priorf{\point}}{\ev}.
\end{split}
\end{align}
The evidence, \ev, is merely a normalization constant in this instance. The likelihood, \like, updates our prior belief with experimental data, resulting in our posterior. 

The posterior in \refeq{Eq:Posterior} is a PDF of all of the model's parameters. To find the PDF for \eg two of the model's parameters, we \ic{marginalize} the posterior;
\beq
\pg{x_1, x_2}{\model,\,\data} = \int \pg{\point}{\model,\,\data}\,\text{d}{x_3}\text{d}{x_4}\cdots
\eeq
Marginalization incorporates \ic{fine-tuning.} If for fixed $(x_1,\,x_2)$, few combinations of $x_3,\,x_4,\ldots$ result in an appreciable posterior density,  \pg{\point}{\model,\,\data}, the marginalized posterior at $(x_1,\,x_2)$ will be small. For further details, see \eg \refcite{Fowlie:2014awa}.

\section{\label{Section:Method}Methodology}
Our calculation of the evidence from \refeq{Eqn:Evidence} requires two ingredients: our priors, which contain our prior beliefs about the model's parameter space, and our likelihood function, which contains relevant experimental data. We supply our priors and our likelihood functions to the nested sampling algorithm with importance sampling implemented in \texttt{(Py)MultiNest-3.4}\cite{Feroz:2008xx,Buchner:2014nha}, which returns the models' Bayesian evidences and posterior PDFs.

This investigation is similar to that in \refcite{Allanach:2006jc,Allanach:2007qk,Cabrera:2008tj,Cabrera:2009dm,Cabrera:2012vu,Fowlie:2014xha}, in which the posterior PDF is calculated for the CMSSM with naturalness priors, \refcite{LopezFogliani:2009np,Kowalska:2012gs}, in which the posterior PDF is calculated for the CNMSSM but without naturalness priors, and \refcite{Kim:2013uxa}, in which a naturalness prior is calculated for the CNMSSM. The posterior PDF for the CNMSSM with naturalness priors and the Bayes-factor for the CNMSSM versus the CMSSM are, however, absent in the literature.

\subsection{\label{sec:like}Likelihood function}
Our likelihood function includes data from relevant laboratory measurements:
\begin{enumerate}[label={(\arabic*)}]

 \item The measured $Z$-boson mass, $\mz=91.1876\gev$\cite{Beringer:1900zz}, with a Dirac likelihood function.
 
 \item Measurements and searches for Higgs bosons at LEP, the Tevatron and the LHC with a $2\gev$ theoretical uncertainty in the Higgs mass\cite{Allanach:2004rh}. The likelihood is from \texttt{HiggsSignals-1.2.0}\cite{Bechtle:2013xfa,Bechtle:2008jh,Bechtle:2011sb,Bechtle:2013gu,Bechtle:2013wla} with the \ic{latest results} dataset (see Fig.~2 of \refcite{Bechtle:2013xfa} for a summary of the experimental data).
 
 \texttt{HiggsSignals-1.2.0} confronts the whole Higgs sector --- all Higgs bosons --- with data. We do not need to separately consider different interpretations of the SM-like Higgs boson in the CNMSSM.

 \item The \texttt{ATLAS-CONF-2013-047}\cite{TheATLAScollaboration:2013fha} 0 leptons + 2-6 jets + MET search with a hard-cut on the \pmm $95\%$ confidence limit. The \pmm confidence limit is approximately independent of \pat\cite{Allanach:2011ut,Bechtle:2011dm,Fowlie:2012im} and the extra CNMSSM singlet superfield\cite{Kowalska:2012gs}.
  
 \item The magnetic moment of the muon calculated with \texttt{SuperIso-3.3}\cite{Mahmoudi:2007vz,Mahmoudi:2008tp,Mahmoudi:2009zz}.

 \item $B$-physics rare decays --- \bsmm, \bsg and \btn~ --- calculated with \texttt{SuperIso-3.3}.

\end{enumerate}
For the numerical values of the constraints, see \reftable{Table:Data}. We calculate mass spectra for the CMSSM and CNMSSM with \texttt{SOFTSUSY-3.4.1}\cite{Allanach:2001kg,Allanach:2013kza}. Our codes for the CNMSSM are consistent with our codes for the CMSSM. Unfortunately, the precise Higgs mass calculation in \texttt{FeynHiggs-2.10.0}\cite{Heinemeyer:1998yj,Heinemeyer:1998np,Degrassi:2002fi,Frank:2006yh} is unavailable in the CNMSSM. We omit observables that we cannot consistently calculate, \eg EW precision observables. 

We exclude DM experiments from our likelihood, \eg the Planck measurement of the DM density from the cosmic microwave background (CMB)\cite{Ade:2013zuv} and the LUX search for DM with an underground detector\cite{Akerib:2013tjd}, because fine-tuning related to DM might cloud our understanding of the fine-tuning of the EW scale. If we were to include DM experiments, we would invoke particular DM annihilation mechanisms by fine-tuning the supersymmetric particle (sparticle) masses and the mixing angles between mass and gauge eigenstates. Including DM experiments would, furthermore, require additional assumptions and uncertainties \see{Baer:2012uya}.  

\begin{table}[ht]
\centering
\begin{ruledtabular}
\begin{tabular}{ccc}
Quantity & Experimental data, $\mu\pm\sigma$ & Theory error, $\tau$\\
\hline
\mz & $91.1876\gev$\cite{Beringer:1900zz}\\
\damu & $(28.8\pm8.0) \times10^{-10}$\cite{Beringer:1900zz} &  $1.0\times10^{-10}$\cite{Heinemeyer:2004gx}\\
\bsmm & $(3.2\pm1.5)\times10^{-9}$\cite{Beringer:1900zz} & $14\%$\cite{deAustri:2006pe}\\
\bsg & $(3.43\pm0.22)\times10^{-4}$\cite{Amhis:2012bh} & $0.21\times10^{-4}$\cite{Misiak:2006zs}\\
\btn & $(1.14\pm0.22)\times10^{-4}$\cite{Amhis:2012bh} & $0.38\times10^{-4}$\cite{Trotta:2008bp}\\
\hline 
\multicolumn{2}{l}{\texttt{ATLAS-CONF-2013-047}\cite{TheATLAScollaboration:2013fha} search for SUSY in $\sim20\invfb$ at \roots{8}.}\\
\multicolumn{2}{l}{LHC, Tevatron and LEP Higgs searches. See Fig.~2 of \refcite{Bechtle:2013xfa}.}\\
\end{tabular}
\end{ruledtabular}
\caption{Experimental data included in our likelihood function.}
\label{Table:Data}
\end{table}

\subsection{Priors}
We pick \ic{naturalness priors}\cite{Allanach:2006jc,Allanach:2007qk,Cabrera:2008tj,Cabrera:2009dm,Cabrera:2012vu,Fichet:2012sn,Kim:2013uxa,Fowlie:2014xha} for the model parameters. That is, we pick priors for the model parameters in the soft-breaking Lagrangian and superpotential at the GUT scale and transform to parameters at the EW scale \eg \tanb, obtained after EW symmetry breaking, with the appropriate Jacobian. 

Traditionally, fine-tuning of the EW scale is measured with partial derivatives of the EW scale with respect to Lagrangian parameters, \eg the Barbieri-Giudice-Ellis measure\cite{Barbieri:1987fn,Ellis:1986yg}. As discussed in \eg \refcite{Fowlie:2014xha}, traditional fine-tuning measures of the EW scale approximate naturalness priors; however, traditional fine-tuning measures lack a probabilistic meaning.

Naturalness priors are an \ic{honest} prior choice. The $(\mz,\,\tanb)$ parameters are output from the fundamental Lagrangian parameters. We are not ignorant of their origin. Our priors ought to reflect that. Typical Bayesian analyses in the literature, \eg \refcite{Kowalska:2012gs,Roszkowski:2014wqa}, pick a linear prior for \tanb and no explicit prior for $\mu$. The implicit prior for $\mu$ in such analyses is that $\mu$ is always such that $\mz=91.1876\gev$\cite{Beringer:1900zz}, \ie
\beq
\priorf{\mu} \propto \delta\(\mu - \mu_Z(\mzero, \mhalf, \azero, \tanb, \ldots)\),
\eeq
where $\mu_Z$ is the numerical value of $\mu$ resulting in the experimentally measured value of \mz for particular input parameters. This is a \ic{dishonest,} informative prior choice.

We pick logarithmic priors for the models' soft-breaking and superpotential parameters, because we are ignorant of their scale, but transform to \tanb and \mz with an appropriate Jacobian. Working with $(\mz,\,\tanb)$ as our input parameters, we are guaranteed to find points with the correct EW scale. The Jacobian in the CMSSM results from trading $(\mu^2,\,b)\to(\mz,\,\tanb)$;
\beq
\mathcal{J}^\text{CMSSM} = \dpartial{\mu^2}{\mz}\dpartial{b}{\tanb} - \dpartial{b}{\mz}\dpartial{\mu^2}{\tanb} = \dpartial{\mu^2}{\mz}\dpartial{b}{\tanb}.
\eeq
The sign of the $\mu$-parameter, \sgnmu, is a discrete input parameter.

In the CNMSSM, we trade $(\msinglet^2,\,\kappa)\to(\mz,\,\tanb)$ resulting in the Jacobian
\beq
\mathcal{J}^\text{CNMSSM} = \dpartial{\kappa}{\mz}\dpartial{\msinglet^2}{\tanb} - \dpartial{\msinglet^2}{\mz}\dpartial{\kappa}{\tanb}. 
\eeq
In addition, we trade $\text{sign}\,\lambda\to\sgnmueff$. The sign of the singlet VEV, $\langle S \rangle$, is unphysical and can be chosen to be always positive, such that $\text{sign}\,\lambda=\sgnmueff$. This transformation, traditional in the literature, simply renames a parameter; there is no associated Jacobian.

We find our naturalness priors by recognizing that if $\priorf{\vec x}$ is a PDF, then
\beq
\priorf{\vec f(\vec x)} = \priorf{\vec x} \times \mathcal{J} \qquad\text{where}\qquad \mathcal{J} = \left|\det \dpartial{x_i}{f_j}\right|.
\eeq

Our naturalness priors for $(\mz,\,\tanb)$ in the CMSSM are
\beq
\label{Eq:MSSM_prior}
\priorf{\mz,\tanb} = \priorf{\mu^2,b} \times \mathcal{J}^\text{CMSSM} \propto \frac{1}{b\mu^2} \times \dpartial{\mu^2}{\mz}\dpartial{b}{\tanb}.
\eeq
Similarly, our naturalness priors for $(\mz,\,\tanb)$ in the CNMSSM are
\beq
\label{Eq:NMSSM_prior}
\priorf{\mz,\tanb} = \priorf{\msinglet^2,\kappa} \times \mathcal{J}^\text{CNMSSM} \propto \frac{1}{\msinglet^2\kappa} \times \mathcal{J}^\text{CNMSSM}.
\eeq
We implement such priors by scanning the models in their $(\mz,\,\tanb)$ parameterizations with naturalness priors. We calculate the Jacobians with numerical differentiation by modifying \texttt{SOFTSUSY-3.4.1} and \texttt{NMSSMSpec-4.2.1}. The naturalness priors for the CNMSSM were recently studied in \refcite{Kim:2013uxa}.

Our prior ranges are in \reftable{Table:Priors}. We pick SUSY breaking masses less than $20\tev$; \refcite{Fowlie:2014awa,Kowalska:2013hha} indicate that the posterior PDF and evidence beyond $20\tev$ is insignificant. If one wishes to enlarge our priors for the SUSY breaking masses beyond $20\tev$, the evidence can be scaled to correct the denominator in the evidence calculation in \refeq{Eqn:Ev_Long} \see{Fowlie:2014xha};
\beq
\ev(\text{Enlarged priors}) = \ev(\text{Priors with $\msusy\le20\tev$}) \times \frac{\text{Volume with $\msusy\le20\tev$}}{\text{Volume of enlarged priors}}.
\eeq
Because the Bayes-factor is a ratio of evidences, this correction cancels for the CMSSM versus the CNMSSM.

We pick the CMSSM $\mu$-parameter less than the Planck scale. By permitting $\mu\gg\msusy$, we incorporate the $\mu$-problem in our analysis. In the CNMSSM, the effective $\mu$-parameter is a function of the SUSY breaking scale. In the CMSSM, the $\mu$-parameter could be far from the SUSY breaking scale. If we picked $\mu\sim\msusy$ in our priors for the CMSSM, we would hide the $\mu$-problem.

We assign zero prior probability to \ic{unphysical} points, \eg points that result in incorrect EW symmetry breaking, an LSP which is not the lightest neutralino, or a Landau pole below the GUT scale. In the CNMSSM, we minimize the occurrence of Landau poles below the GUT scale in $\lambda$ by choosing $\lambda\le4\pi$ at the GUT scale in our priors in \reftable{Table:Priors}.

\begin{table}[ht]
\centering
\begin{ruledtabular}
\begin{tabular}{cc}
Parameter & Distribution\\
\hline
CMSSM\\
\hline
\mzero & Log, $0.3\dash20\tev$\\
\mhalf & Log, $0.3\dash10\tev$\\
\azero & Flat, $-20\dash20\tev$\\
$\mu$ & Log, $1\gev\dash\mplanck$\\
$b$ & Log, $0.3\dash20\tev$\\
\sgnmu & $\pm1$ with equal probability\\
\hline
CNMSSM\\
\hline
\mzero & Log, $0.3\dash20\tev$\\
\mhalf & Log, $0.3\dash10\tev$\\
\azero & Flat, $-20\dash20\tev$\\
$\lambda$ & Log, $0.001\dash4\pi$\\
\msinglet & Log, $0.3\dash20\tev$\\
$\kappa$  & Log, $0.001\dash4\pi$\\
\sgnmueff & $\pm1$ with equal probability\\
\hline
SM\\
\hline
\mb       & Gaussian, $4.18\pm0.03\gev$\cite{Beringer:1900zz}\\
\mt       & Gaussian, $173.07\pm0.89\gev$\cite{Beringer:1900zz}\\
\invalpha & Gaussian, $127.944\pm0.014$\cite{Beringer:1900zz}\\
\alphas   & Gaussian, $0.1196\pm0.0017$\cite{Beringer:1900zz}\\
\hline
Phenomenological\\
\hline
\tanb  & Effective, \refeq{Eq:MSSM_prior}, $2\dash62$\\
\mz    & Effective, \refeq{Eq:MSSM_prior}, $91.1876\gev$\cite{Beringer:1900zz}\\
\end{tabular}
\end{ruledtabular}
\caption{Priors for the CMSSM and CNMSSM model parameters.}
\label{Table:Priors}
\end{table}

\subsection{Evidence}
Let us clarify the calculation of the Bayesian evidence. In the CMSSM, we wish to calculate the evidence by picking priors in the $(\mu^2,\,b)$ parameterization;
\beq
\label{Eqn:Ev_Long}
\ev = \frac{\int_{R(\mu^2,\,b)}\text{d}\mu^2\text{d}b\,\int\text{d}\cdots\,\likef{\mu^2,b,\cdots} \times\priorf{\mu^2,b\cdots}}
{\int_{R(\mu^2,\,b)}\text{d}\mu^2\text{d}b\,\int\text{d}\cdots\,\priorf{\mu^2,b,\cdots}}.
\eeq
The ellipses represent the model's other parameters. The priors, \prior, are unnormalized, hence the denominator. The region of integration, $R(\mu^2,\,b)$, is the prior ranges in \reftable{Table:Priors}.

We could compute the integral in the numerator of \refeq{Eqn:Ev_Long} with Monte Carlo (MC) integration; however, because few points would predict the correct EW scale, finding modes in the likelihood function would be time-consuming. If we change variables to $(\mz,\,\tanb)$, we guarantee that points predict the correct EW scale;
\beq
\label{Eqn:Ev_Calc}
\ev = \frac{\int_{R(\mz,\,\tanb)}\text{d}\mz\text{d}\tanb\,\int\text{d}\cdots\,\likef{\mz,\tanb,\cdots}\times\priorf{\mu^2,b,\cdots}\times\mathcal{J}}
{\int_{R(\mu^2,\,b)}\text{d}\mu^2\text{d}b\,\int\text{d}\cdots\,\priorf{\mu^2,b,\cdots}}.
\eeq
The change of variables introduces the Jacobian that we calculate for our naturalness priors. For the change in the integration region to $R(\mz,\,\tanb)$, we make an approximation. We pick $R(\mz,\,\tanb)$ to be the region in $(\mz,\,\tanb)$ in which the likelihood is appreciable. The regions in $(\mz,\,\tanb)$ in which the likelihood is not appreciable cannot significantly contribute to the integral. We trust that the original $R(\mu^2,\,b)$ region spans at least that region in $(\mz,\,\tanb)$.

For reproducibility, we note that if one includes naturalness priors in a \ic{likelihood} supplied to \texttt{(Py)MultiNest-3.4}, it returns
\beq
\label{Eqn:Ev_MN}
\ev^\prime = \frac{\int_{R(\mz,\,\tanb)}\text{d}\mz\text{d}\tanb\,\int\text{d}\cdots\,\likef{\mz,\tanb,\cdots}\times\priorf{b,\mu,\cdots}\times\mathcal{J}}
{\int_{R(\mz,\,\tanb)}\text{d}\mz\text{d}\tanb\,\int\text{d}\cdots\,\priorf{\cdots}},
\eeq
\ie without a Jacobian in the denominator. The difference between \refeq{Eqn:Ev_Calc} and \refeq{Eqn:Ev_MN} must be corrected by hand;
\beq
\ev = \ev^\prime \times \frac{%
\int_{R(\mz,\,\tanb)}\text{d}\mz\text{d}\tanb%
}{%
\int_{R(\mu^2,b)}\text{d}\mu^2\text{d}b\,\priorf{\mu^2,b}%
}.
\eeq
In the CNMSSM, our calculation is similar.

\section{\label{Section:Results}Results}
\subsection{Posterior}
We inspect the posterior in the CMSSM and CNMSSM by plotting $1\sigma$ and $2\sigma$ credible regions on marginalized two-dimensional planes. Our $1\sigma$ and $2\sigma$ credible regions are the smallest regions that contain $68\%$ and $95\%$ of the posterior; the regions in which the posterior is most dense. One can always draw credible regions; the existence and size of the credible regions is not indicative of agreement with data or the absence of fine-tuning. 

\begin{figure}[t]
\centering
\subfloat[][CMSSM.]{\label{fig:m0m12:i}
\includegraphics[width=0.49\linewidth]{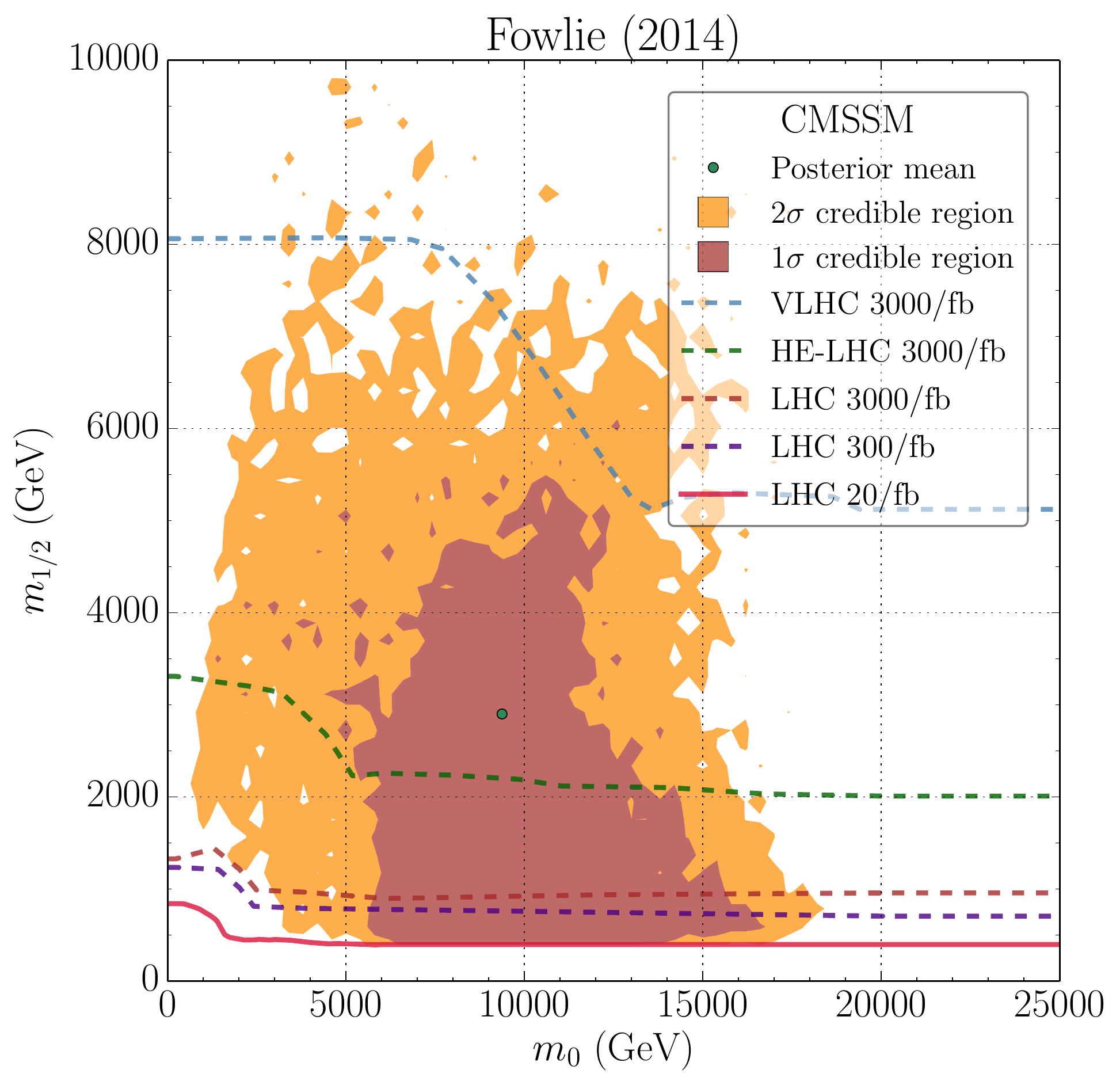}
}
\subfloat[][CNMSSM.]{\label{fig:m0m12:ii}
\includegraphics[width=0.49\linewidth]{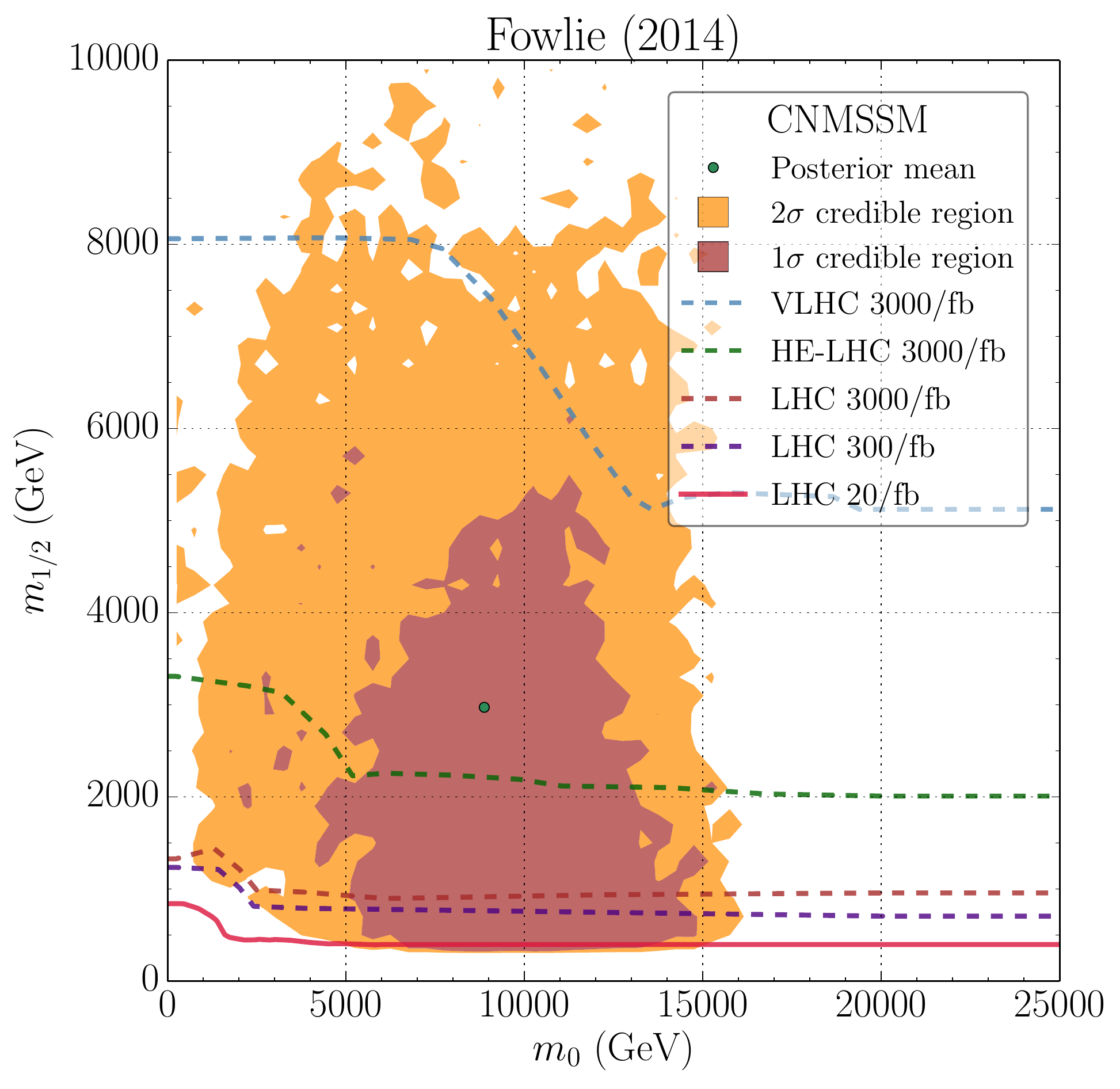}
}
\caption{The \pmm planes of the \protect\subref{fig:m0m12:i} CMSSM and  \protect\subref{fig:m0m12:ii} CNMSSM showing the $68\%$ (red) and $95\%$ (orange) credible regions of the marginalized posterior. The $95\%$ exclusion from \texttt{ATLAS-CONF-2013-047}\cite{TheATLAScollaboration:2013fha} is shown with a solid line. The expected discovery reaches of future hadron colliders from \refcite{Fowlie:2014awa} are also shown with dashed lines.}
\label{fig:m0m12}
\end{figure}

Let us compare the CMSSM and CNMSSM side-by-side, beginning with their \pmm planes in \reffig{fig:m0m12}. The posterior favors gaugino masses as light as is permitted by the exclusion contour from the LHC, approximately $\mhalf\gtrsim0.5\tev$; however, the $1\sigma$ credible region extends to $\mhalf\lesssim6\tev$. The $1\sigma$ credible region for the unified scalar mass spans $5\tev\lesssim\mzero\lesssim15\tev$. The difference between the CNMSSM's and the CMSSM's \pmm planes is small; in the CNMSSM, \mhalf is slightly larger and \mzero is slightly smaller than that in the CMSSM. 

\latin{Prima facie,} that  $\mzero\gtrsim5\tev$ is surprising; scalar masses closer to the EW scale, in \eg the \stauc\cite{Ellis:1998kh} and $A$-funnel\cite{Drees:1992am} DM annihilation regions, are permitted by the likelihoods, but excluded by the posterior. The discovery reaches in \reffig{fig:m0m12} from \refcite{Fowlie:2014awa} indicate that the \roots{14} LHC and a \roots{33} High-Energy LHC (HE-LHC) might struggle to discover the CMSSM or CNMSSM, but that a \roots{100} Very Large Hadron Collider (VLHC) would probably discover the CMSSM or CNMSSM were nature described by either model.

The posterior favors $5\tev\lesssim\mzero\lesssim15\tev$ because of \ic{focusing} in the renormalization group (RG) equations for the soft-breaking masses\cite{Feng:1999zg,Feng:2000bp,Feng:2011aa}. With focusing in the RG equations, the up-type soft-breaking Higgs mass at the EW scale is similar to the EW scale,
\beq
\mhu|_{\text{EW}} \sim \mweak,
\eeq
and is approximately independent of the initial values of the soft-breaking masses at the GUT scale.\footnote{Focusing is not, however, a fixed point in the RG flow \see{Feng:2000bp}.} The RG running of \eg squark and slepton soft-breaking masses is not focused to the EW scale; the squarks and sleptons could be much heavier than the EW scale. Regions of parameter space in which the up-type Higgs mass is focused  generically predict the correct EW scale via \refeq{Eqn:MZ} without fine-tuning; they are natural. The modes in the posterior at $5\tev\lesssim\mzero\lesssim15\tev$ in \reffig{fig:m0m12} are \ic{focus points.}

\begin{figure}[t]
\centering
\subfloat[][CMSSM.]{\label{fig:a0tanb:i}
\includegraphics[width=0.49\linewidth]{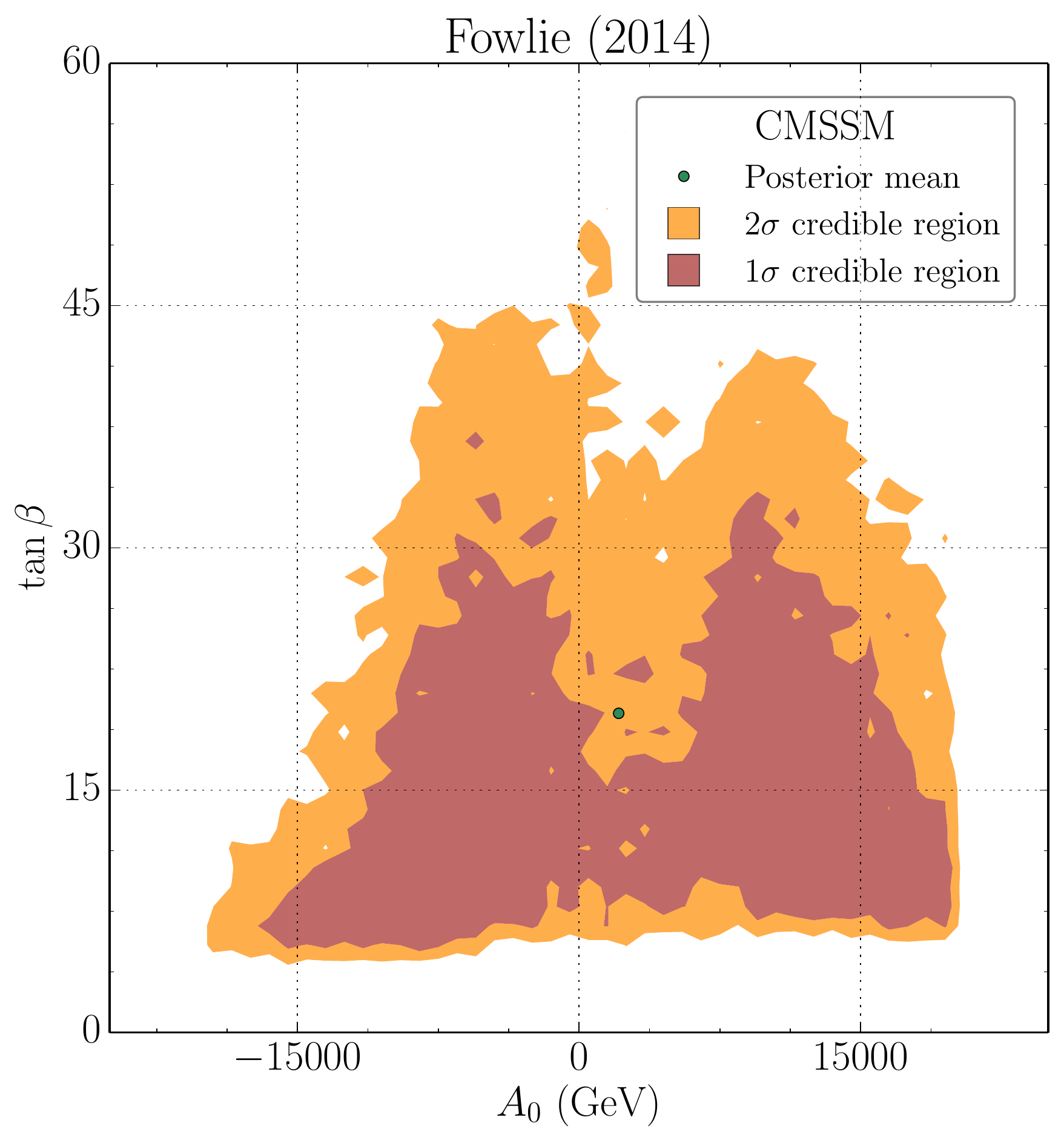}
}
\subfloat[][CNMSSM.]{\label{fig:a0tanb:ii}
\includegraphics[width=0.49\linewidth]{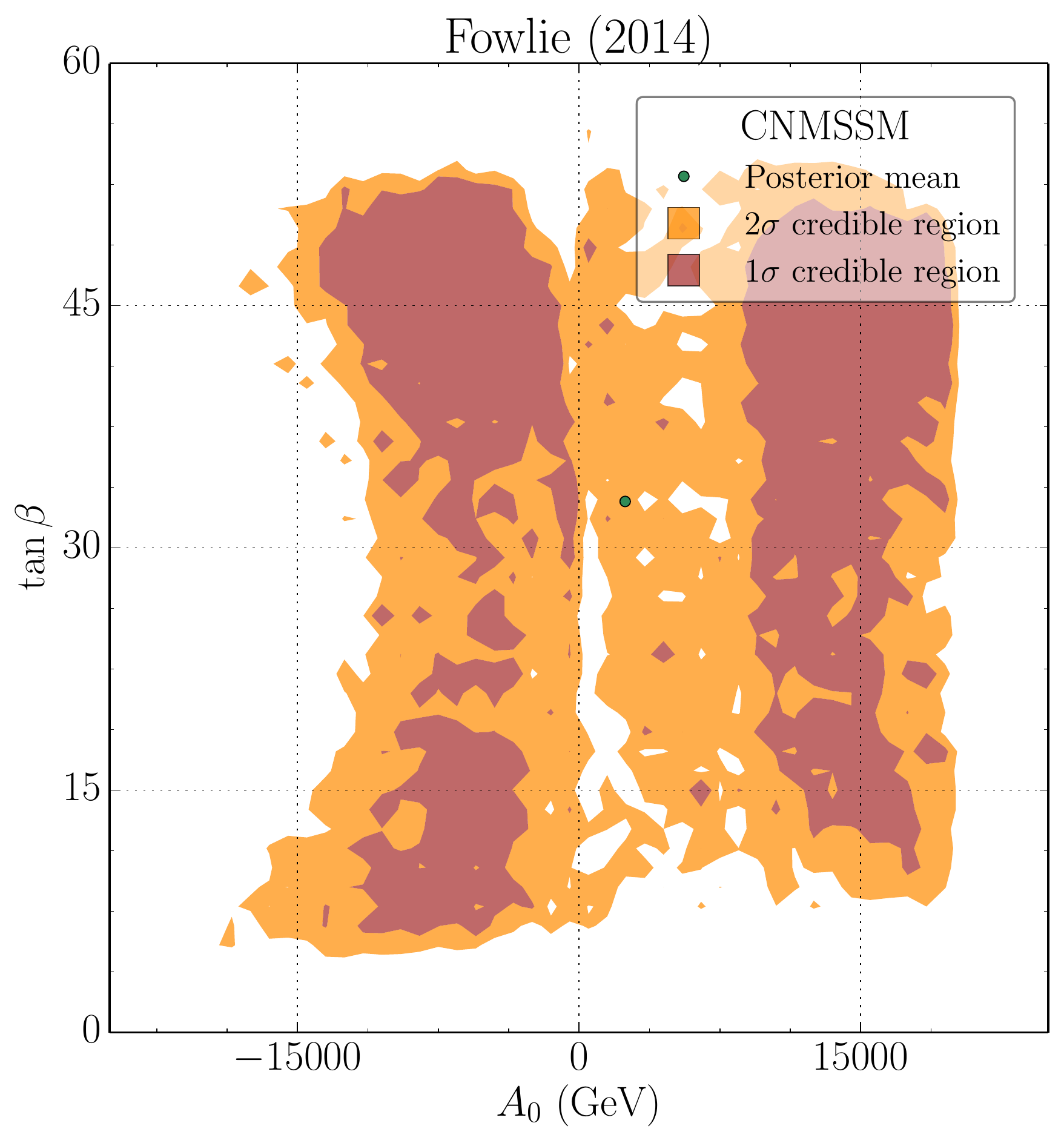}
}
\caption{The \pat planes of the \protect\subref{fig:a0tanb:i} CMSSM and  \protect\subref{fig:a0tanb:ii} CNMSSM showing the $68\%$ (red) and $95\%$ (orange) credible regions of the marginalized posterior.}
\label{fig:a0tanb}
\end{figure}

On the CMSSM's \pat plane in \reffig{fig:a0tanb:i}, the  $1\sigma$ credible region spans a wide range of trilinear, $|\azero|\lesssim20\tev$, but a restricted range of \tanb, $\tanb\lesssim30$ and $\tanb\lesssim15$ if $|\azero|\gtrsim10\tev$. This behavior is expected; large \tanb is unnatural. By the derivatives in the Jacobians in \refeq{Eq:MSSM_prior} and \refeq{Eq:NMSSM_prior}, our naturalness priors disfavor large \tanb. \refcite{Cabrera:2009dm} explains this simply; from EW symmetry breaking conditions, 
\beq
\tanb \simeq \left.\frac{\mhu^2 + \mhd^2 + 2\mu^2}{b}\right|_{\text{EW}}.
\eeq
In the denominator, the radiative corrections to $b$ from the RG flow are proportional to $\mu\msusy$, whereas the numerator is proportional to $\msusy^2$; rearranging,
\beq
\label{Eqn:tanb}
\msusy \sim \mu\tanb.
\eeq
Large \tanb implies a hierarchy between the soft-breaking masses and the $\mu$-parameter. The EW scale, however, results from a cancellation between the soft-breaking masses and the $\mu$-parameter; thus large \tanb implies fine-tuning. Because $b\propto\tanb$, our logarithmic prior for $b$ also disfavors large \tanb. \tanb could, however, enhance focusing by affecting the top and bottom Yukawa couplings.

On the CNMSSM's \pat plane in \reffig{fig:a0tanb:ii}, however, \tanb is larger than in the CMSSM at $1\sigma$ in \reffig{fig:a0tanb:i}, with $\tanb\lesssim50$. We cannot apply our previous argument, that large \tanb is unnatural, to the CNMSSM, because the $\mu$-parameter is a function of the SUSY breaking scale. Because large \tanb in the CNMSSM does not imply a hierarchy between scales or fine-tuning, it is not penalized by our effective priors.

\begin{figure}[t]
\centering
\subfloat[][CMSSM.]{\label{fig:mass:i}
\includegraphics[width=0.49\linewidth]{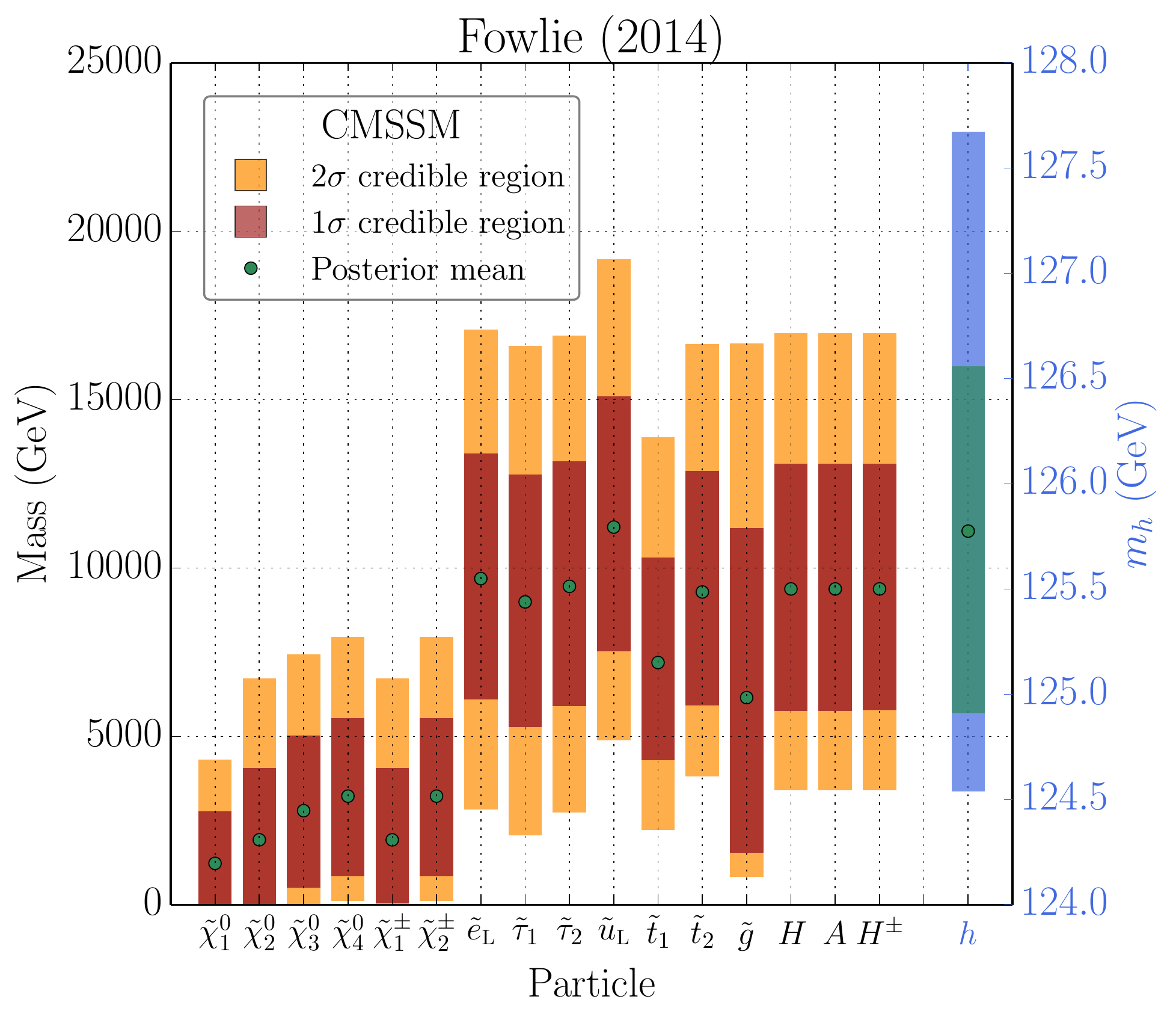}
}
\subfloat[][CNMSSM.]{\label{fig:mass:ii}
\includegraphics[width=0.49\linewidth]{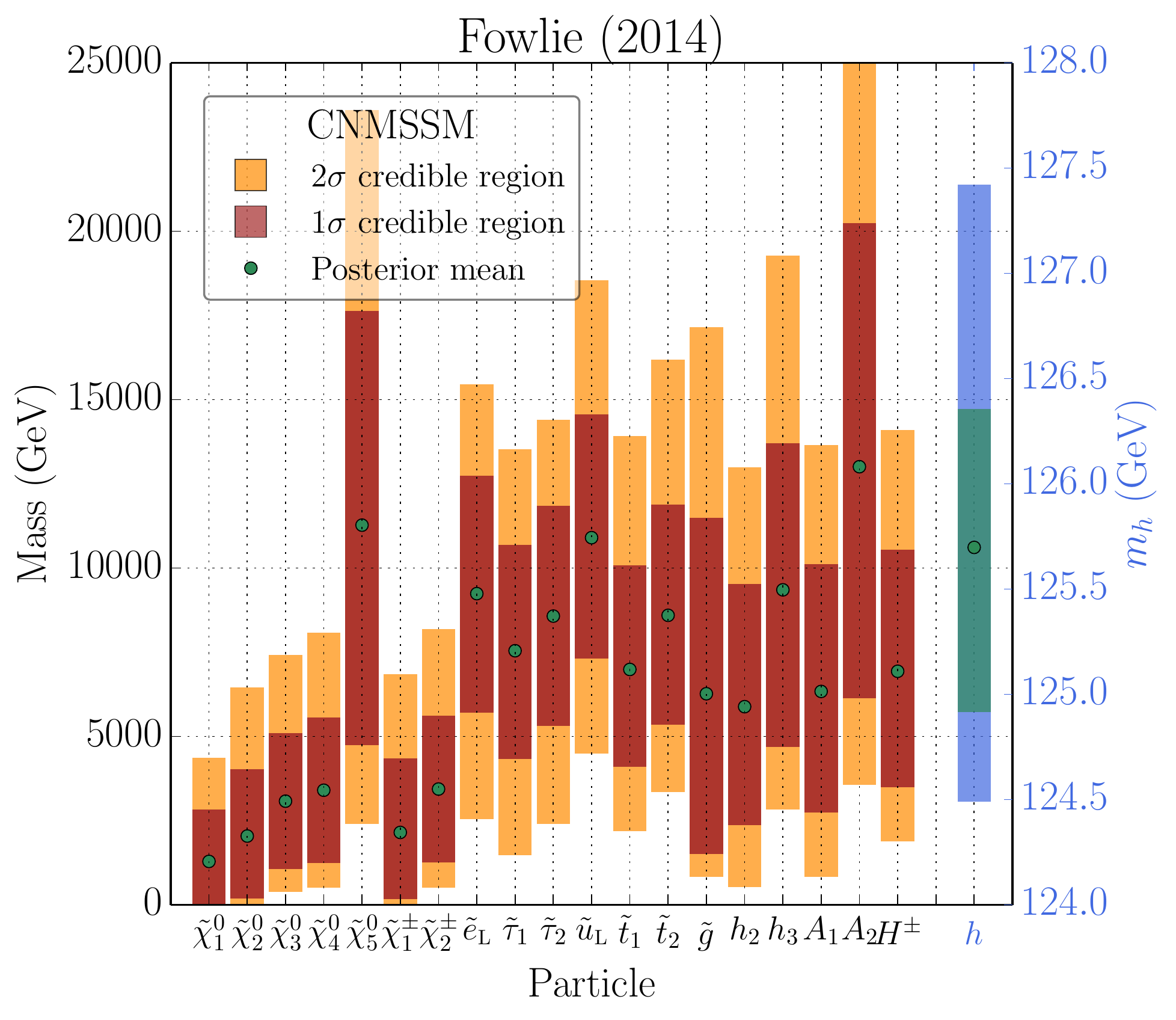}
}
\caption{Sparticle masses in the \protect\subref{fig:mass:i} CMSSM and \protect\subref{fig:mass:ii} CNMSSM.
The red and orange bars are the $68\%$ and $95\%$ credible regions for the sparticle masses. 
The green and blue bars are the $68\%$ and $95\%$ credible regions for the Higgs mass; note that the Higgs mass has a separate scale.
The circles are the posterior means.}
\label{fig:mass}
\end{figure}

We compare the sparticle and Higgs masses in the CMSSM and CNMSSM in \reffig{fig:mass}. In both models, the lightest neutralino is typically Higgsino-like or a mixture of Higgsino and gaugino gauge eigenstates, because $\mu$ is small. The sleptons, squarks and gluino are between approximately $5\tev$ and $15\tev$, though slightly heavier in the CMSSM than in the CNMSSM. With such heavy squarks, the Higgs mass is $\mh\simeq126\gev$, in agreement with experiment. In the CNMSSM, the Higgs with a mass of about $126\gev$ is always the lightest Higgs. As anticipated, $\mh\simeq126\gev$ is achieved in the CNMSSM with slightly lighter sparticles than in the CMSSM, because of the CNMSSM's additional tree-level contribution to the Higgs mass in \refeq{Eqn:N_Higgs}. 

\begin{figure}[t]
\centering
\subfloat[][CMSSM.]{\label{fig:higgs:i}
\includegraphics[width=0.49\linewidth]{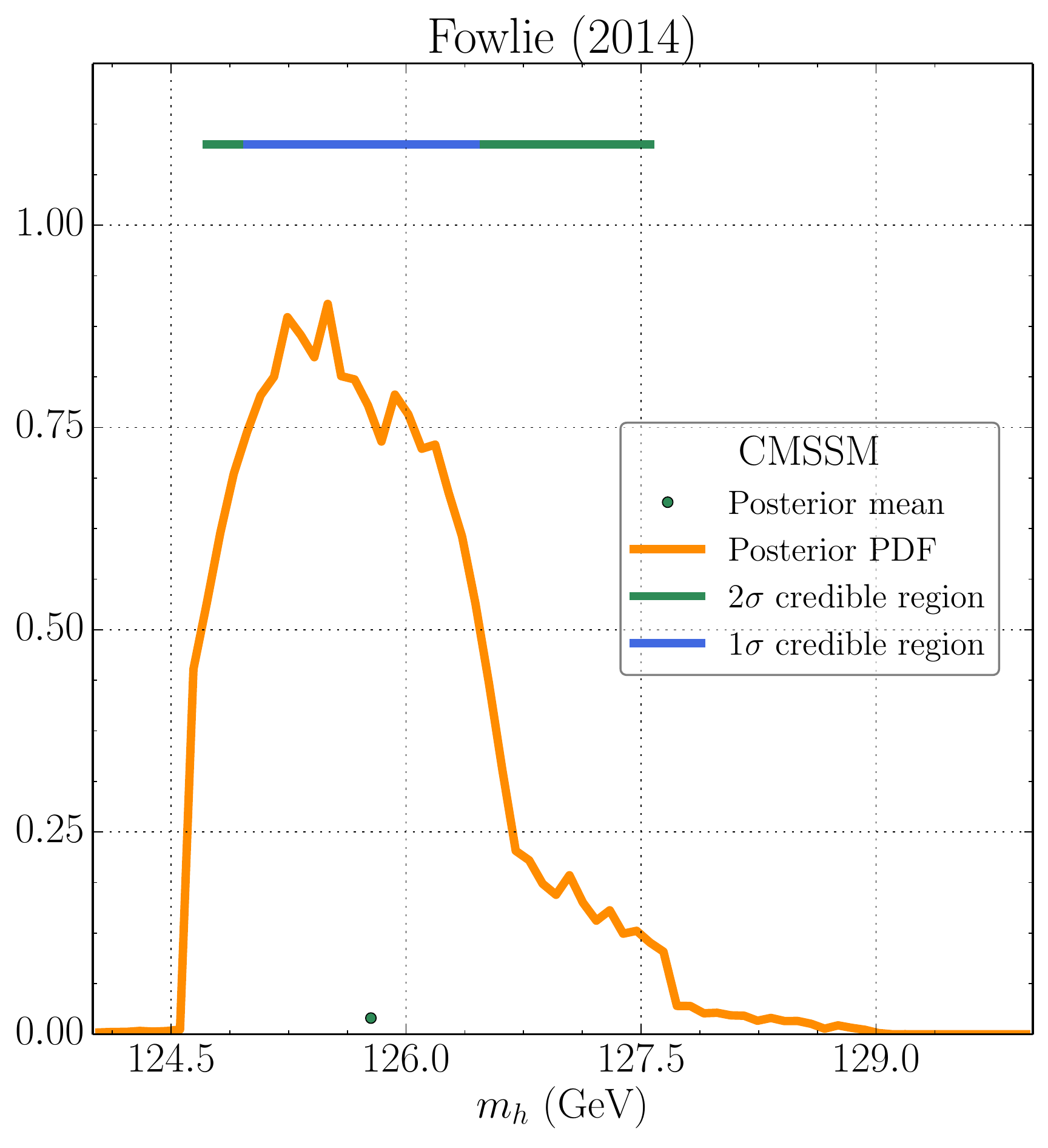}
}
\subfloat[][CNMSSM.]{\label{fig:higgs:ii}
\includegraphics[width=0.49\linewidth]{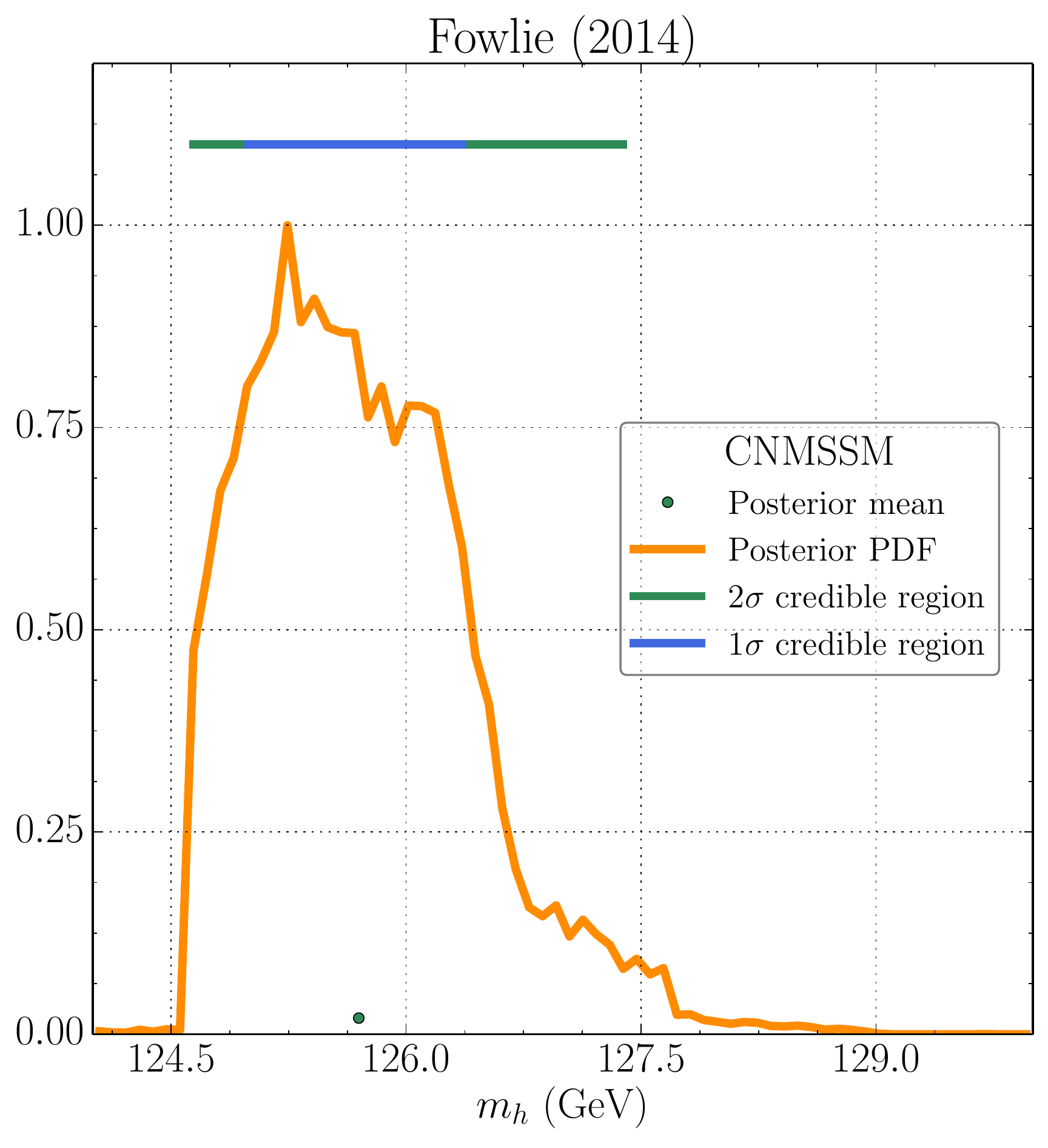}
}
\caption{The predicted Higgs mass in the \protect\subref{fig:higgs:i} CMSSM and \protect\subref{fig:higgs:ii} CNMSSM. The orange line is the marginalized posterior PDF. The green and blue bars are the $68\%$ and $95\%$ credible regions for the Higgs mass. The circles are the posterior means. So that the PDFs can be fairly compared, both PDFs are normalized such that their integrals are identical.}
\label{fig:higgs}
\end{figure}

We further examine the Higgs mass in \reffig{fig:higgs}, in which we plot the one-dimensional PDF for the Higgs mass in the CMSSM and in the CNMSSM. The PDF in the CMSSM and CNMSSM are nearly identical.\footnote{Minor differences in the PDF could result from statistical noise.} Whilst \refeq{Eqn:N_Higgs} indicates that the Higgs mass in the CNMSSM ought to be heavier than that in the CMSSM, the similarity in the PDFs is unsurprising. Our likelihood included a requirement that $\mh\sim126\gev$.

\begin{figure}[t]
\centering
\subfloat[][Additional tree-level contribution to Higgs mass.]{\label{fig:N_dhiggs:i}
\includegraphics[width=0.33\linewidth]{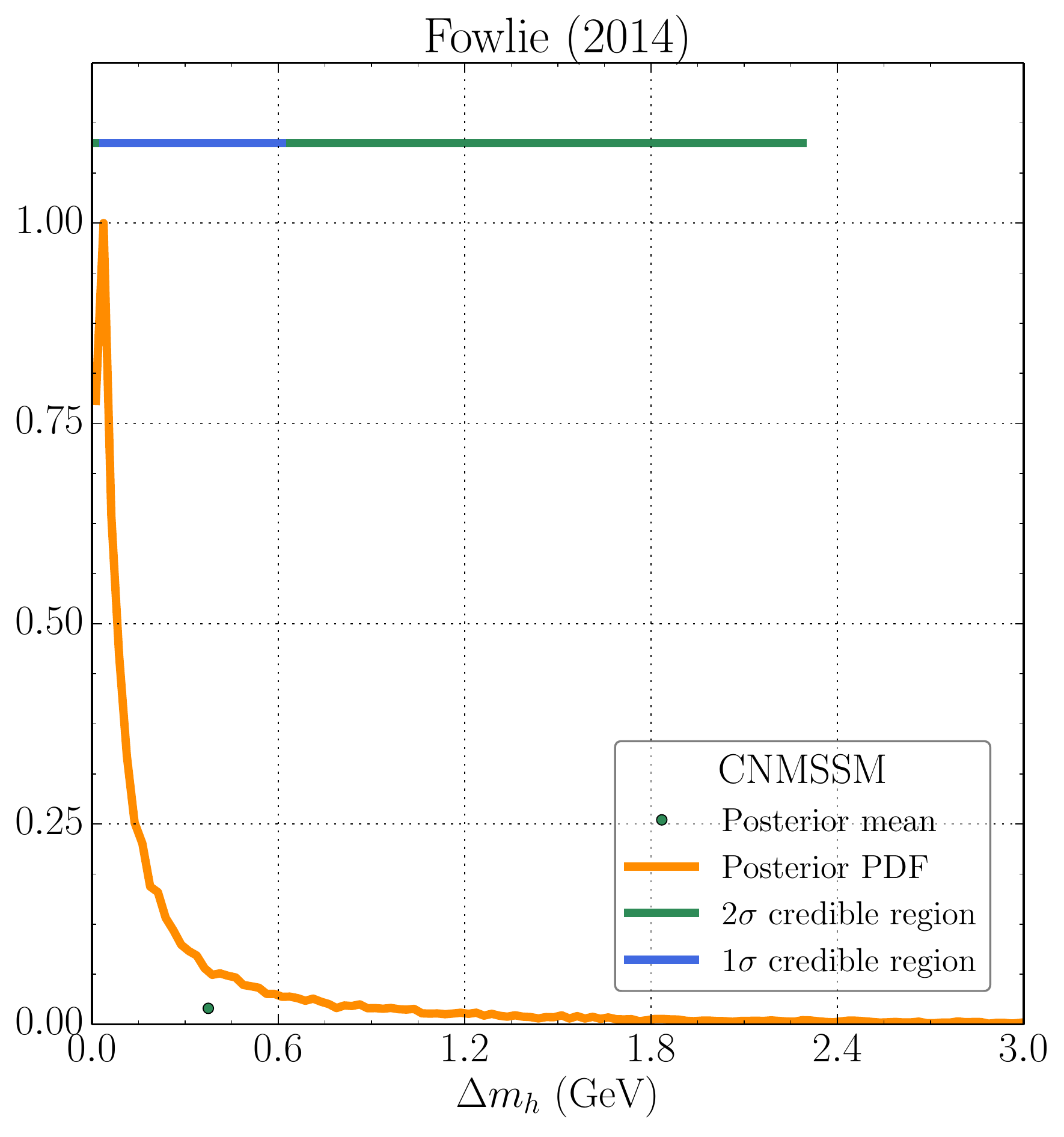}
}
\subfloat[][$(\lambda,\tanb)$ plane.]{\label{fig:N_dhiggs:ii}
\includegraphics[width=0.33\linewidth]{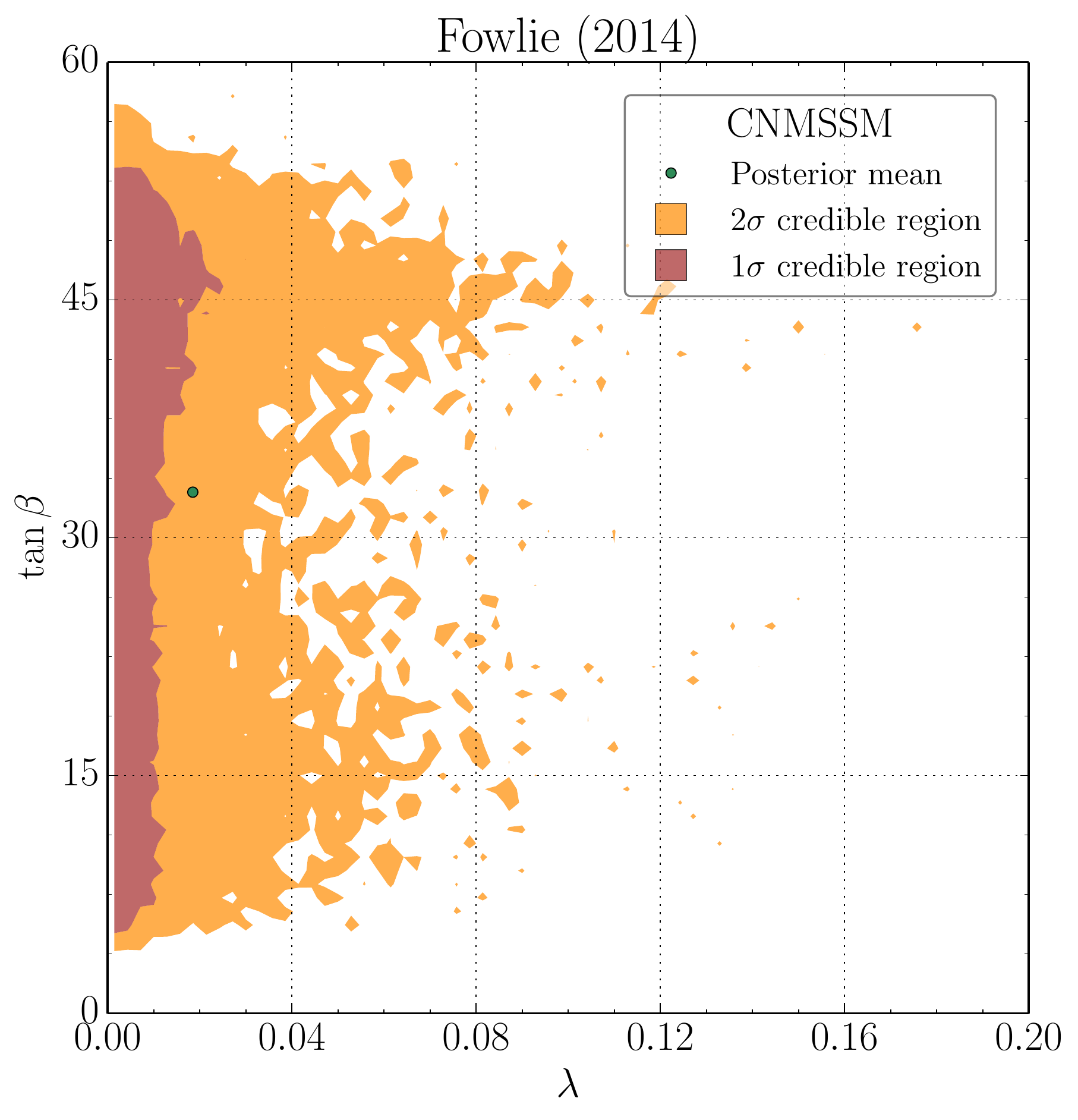}
}
\subfloat[][$(\lambda,\mh)$ plane.]{\label{fig:N_dhiggs:iii}
\includegraphics[width=0.33\linewidth]{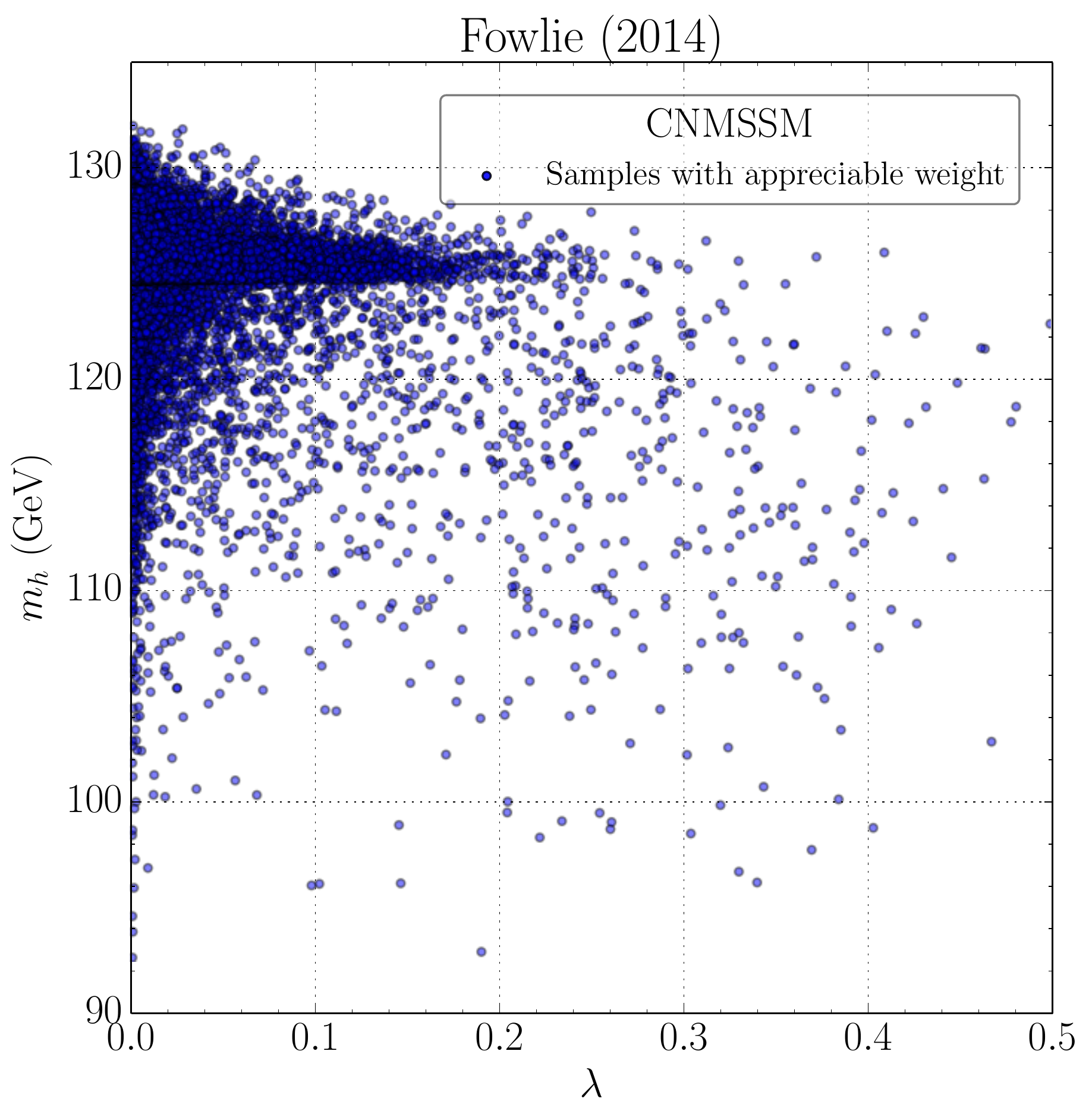}
}
\caption{\protect\subref{fig:N_dhiggs:i} The additional tree-level contribution to Higgs mass in the CNMSSM. The orange line is the marginalized posterior PDF. The green and blue bars are the $68\%$ and $95\%$ credible regions. \protect\subref{fig:N_dhiggs:ii} The $(\lambda,\tanb)$ plane in the CNMSSM showing the $68\%$ (red) and $95\%$ (orange) credible regions of the marginalized posterior. \protect\subref{fig:N_dhiggs:iii} Samples with appreciable posterior weight scattered on the $(\lambda,\mh)$ plane. The relationship between $\lambda$, \tanb and $\Delta m_h$ is in \refeq{Eq:N_dhiggs}.}
\label{fig:N_dhiggs}
\end{figure}

Let us instead examine whether the additional tree-level contribution to the Higgs mass in the CNMSSM in \refeq{Eqn:N_Higgs},
\beq
\label{Eq:N_dhiggs}
\Delta\mh = \lambda v \sin 2\beta,
\eeq
is appreciable. This contribution is added in quadrature, $\mh^2 + \Delta\mh^2$, weakening its impact. We plot $\Delta\mh$ and the relevant parameters, \tanb and $\lambda$, in \reffig{fig:N_dhiggs}. The additional tree-level contribution in the CNMSSM in \reffig{fig:N_dhiggs:i} is negligible; with one tail at $1\sigma$, $\Delta\mh\lesssim0.25\gev$. The smallness of this contribution stems from the smallness of $\lambda$ in \reffig{fig:N_dhiggs:ii}; $\lambda\lesssim0.1$ is favored, although $\lambda$ as large as $4\pi$ is permitted. 

The smallness of $\lambda$ was remarked upon in previous Bayesian studies of the CNMSSM\cite{LopezFogliani:2009np,Kowalska:2012gs,Kim:2013uxa}, in which it was posited that small $\lambda$ minimized the occurrence of tachyonic Higgs bosons. Furthermore, \refcite{Kim:2013uxa} suggests that for $\mh\sim126\gev$ and $\tanb\gtrsim10$, naturalness priors might favor small $\lambda$ (see \eg Fig.~3 in \refcite{Kim:2013uxa}).

\refcite{Ellwanger:2009dp} remarks that if $A_\lambda$ is large, increases in $\lambda$ might decrease the Higgs mass. Because we always select $\mh\sim126\gev$, this behavior is difficult to study; however, \reffig{fig:N_dhiggs:iii}, a scatter plot on the $(\lambda,\mh)$ plane, indicates that this behavior occurs. The highest Higgs mass achieved decreases as $\lambda$ is increased. We caution the reader that the scatter plot is misleading, however, because the density of points cannot be resolved. There are many more points, and much more posterior weight, with $\lambda\lesssim0.1$. In fact, with one tail at $2\sigma$, $\lambda\lesssim0.08$. 

With the Bayesian evidence, fine-tuning is a property of a \ic{neighborhood} in a model's parameter space, \ie the evidence in a \ic{neighborhood} is a probability density multiplied by a volume element. By itself, a probability density is not a well-defined property of an individual point, because it is not \eg invariant under reparameterizations. \refcite{Kowalska:2014hza,Mustafayev:2014lqa,Baer:2014ica} present individual points with small fine-tuning measures. We refrain from presenting such points, because they have no particular probabilistic meaning. 

\subsection{Evidence}
Let us recapitulate our aim. We wanted to find the Bayes-factor for the CNMSSM versus the CMSSM. The Bayes-factor measures how our relative belief in the CNMSSM versus the CMSSM ought to change in light of the experimental data. Bayesian naturalness of the EW scale is automatically incorporated in the Bayes-factor. We interpret the Bayes-factor with the Jeffreys' scale in \reftable{Table:Jeffreys_Scale}. 

If the Bayes-factor is greater than (less than) one, the CNMSSM (CMSSM) is favored. The Bayes-factor was
\beq
B\(\text{CNMSSM}/\text{CMSSM}\) = 10^{+100}_{-5}.
\eeq
The large uncertainty results from the evidence calculation in the CNMSSM. With a reasonable computer time, \texttt{(Py)MultiNest-3.4} found the CNMSSM's evidence with an upper bound one order of magnitude greater than its estimate and the CMSSM's evidence to within a factor of one half. These uncertainties could be reduced with extensive computing resources. Fortunately, the uncertainty in the Bayes-factor corresponds to an uncertainty of a single grade on the Jeffreys' scale in \reftable{Table:Jeffreys_Scale}. The Bayes-factor is \ic{positive} or \ic{strong} evidence in favor of the CNMSSM versus the CMSSM. \ic{Positive} evidence is two grades below \ic{very strong} evidence and one grade above \ic{barely worth mentioning.} 

A factor of about $5$ in this ratio, however, resulted from the difference in the prior volume of $\mu$ in the CMSSM and $\kappa$ in the CNMSSM in \reftable{Table:Priors};
\beq
\frac{\ln\left(\frac{\mplanck}{1\gev}\right)}{\ln\left(\frac{4\pi}{0.001}\right)} \approx 5.
\eeq
This factor is related to the $\mu$-problem \see{Fowlie:2014xha}. Without this factor, the evidence in favor of the CNMSSM versus the CMSSM is \ic{barely worth mentioning} or \ic{positive.} The naturalness of the CNMSSM is overstated in the literature. The difference in the credibility of the CNMSSM and CMSSM is \ic{barely worth mentioning} or \ic{positive,} unless one considers the $\mu$-problem. If one ignores the $\mu$-problem, the evidence in favor of the CNMSSM is unlikely to be \ic{strong} and is certainly not \ic{very strong.}

We anticipated that the CNMSSM would be more credible than the CMSSM, because additional tree-level contributions to the Higgs mass in \refeq{Eqn:N_Higgs} might permit lighter stops. Whilst the stops in the CNMSSM were slightly lighter than the stops in the CMSSM, the stops were $3\tev\lesssim m_{\s{t}} \lesssim 15\tev$ in each model (see \reffig{fig:mass}). We found \ic{barely worth mentioning} to \ic{positive} evidence that the agreement between generic predictions and experimental data in the CNMSSM is better than that in the CMSSM, if one ignores the $\mu$-problem, and \ic{positive} to \ic{strong} evidence if one considers the $\mu$-problem.

The final step, which we omit, is multiplying the Bayes-factor by one's prior odds to find one's relative belief in the CNMSSM versus the CMSSM, in light of experimental data, \ie the posterior odds in \refeq{Eqn:Odds}. When picking prior odds, one must discard knowledge of the EW scale, all experimental data, the fact that the CNMSSM solves the $\mu$-problem of the CMSSM, and any other naturalness considerations that originate from knowledge of the EW scale. To include such knowledge in one's prior odds would be \ic{double-counting;} it is already included in the Bayes-factor.


Calculating Bayesian evidences is numerically challenging and we acknowledge that our evidences suffered from substantial uncertainties. To help judge those uncertainties, we list our \texttt{MultiNest-3.4} settings in \reftable{Table:MN_Settings}. We followed the recommendations in \refcite{Feroz:2011bj} for an accurate calculation of the Bayesian evidence, with the exception of the stopping criteria (the evidence tolerance) in the CNMSSM. Satisfying the stopping criteria recommended in \refcite{Feroz:2011bj} in the CNMSSM would require extensive computing resources. As a consequence, there is an appreciable uncertainty in the evidence in the CNMSSM, as already discussed.

\begin{table}[ht]
\centering
\begin{ruledtabular}
\begin{tabular}{lll}
 & CMSSM & CNMSSM\\
\hline
Samples in posterior distribution & $40\,000$ & $40\,000$\\
Total likelihood evaluations & $400\,000$ & $1\,100\,000$\\
Evidence tolerance & $0.5$ & $8$\\
\hline
\texttt{MultiNest} & \multicolumn{2}{c}{\texttt{v3.4} with \texttt{gcc}}\\
Importance sampling & \multicolumn{2}{c}{True}\\
Multimodal & \multicolumn{2}{c}{False}\\
Constant efficiency & \multicolumn{2}{c}{False}\\
Efficiency & \multicolumn{2}{c}{$1$}\\
Live points & \multicolumn{2}{c}{$4000$}\\
\end{tabular}
\end{ruledtabular}
\caption{Our settings for the \texttt{MultiNest} algorithm. For details, see the \texttt{MultiNest} documentation\cite{Feroz:2008xx}.}
\label{Table:MN_Settings}
\end{table}

\subsection{Possible impact of DM}
As mentioned in \refsec{sec:like}, to avoid extra assumptions and sources of fine-tuning, we omitted DM observables from our likelihood. One might wonder, however, how DM observables might impact our conclusions, were we to assume that the LSP accounted for all of the DM in the Universe. 

In the CNMSSM, because we find that in our results the singlino is decoupled, the singlino is probably irrelevant to DM. We conjecture that DM observables could impact the posterior PDF on the \pmm plane in the CNMSSM and in the CMSSM. Common DM annihilation mechanisms, such as coannihilation or resonances, preclude focusing of the EW scale, and would be disfavored. Focus point regions in which the LSP is a fine-tuned bino-Higgsino mixture could satisfy DM constraints\cite{Feng:2011aa}. Such regions would be favored. 

It is unclear, however, how DM observables might impact the Bayes-factor, \ie whether DM might favor a particular model. Because the singlino is decoupled in the CNMSSM, in each model, the most probable DM is a fine-tuned bino-Higgsino mixture. We find insufficient reason to believe that such a fine-tuned mixture could be more readily achieved in a particular model. As such, we conjecture that the inclusion of DM observables might not significantly impact the Bayes-factor.

\section{\label{Section:Conclusions}Conclusions}
We calculated the posterior PDF and evidence for the CNMSSM and the CMSSM with naturalness priors, including relevant data from the LHC. Previous calculations of the posterior PDF for the CNMSSM picked informative priors for $(\mz,\,\tanb)$ at the EW scale. We picked \ic{honest} priors for the model parameters in the Lagrangian and superpotential at the GUT scale. Whilst such priors were calculated for the CNMSSM in \refcite{Kim:2013uxa}, the posterior PDF and evidence for the CNMSSM with such priors are absent in the literature.

We examined the credible regions of the CMSSM and CNMSSM, finding which regions of parameter space were favored by Bayesian naturalness. Mechanisms that focus Higgs SUSY breaking masses to the EW scale were favored. In each model, the SUSY breaking masses were $\mhalf\lesssim8\tev$ and $\mzero\lesssim15\tev$, with squarks and sleptons $\sim10\tev$. The discovery prospects at the LHC were limited; with $3000\invfb$ of data, it is unlikely that the LHC could discover either the CMSSM or the CNMSSM. Contrariwise, the HE-LHC would probably discover the CMSSM or CNMSSM, were nature described by either model.

We computed the Bayes-factor for the CNMSSM versus the CMSSM. The calculation involved moderate uncertainties that could be resolved with extensive computing resources. We found that the evidence in favor of the CNMSSM versus the CMSSM is \ic{positive} to \ic{strong} on the Jeffreys' scale, but that if one ignores the $\mu$-problem, the evidence is \ic{barely worth mentioning} to \ic{positive.} \ic{Positive} evidence is two grades below \ic{very strong.} We conclude that the credibility of the CNMSSM is perhaps overstated in the literature and that the $\mu$-problem must be considered in a comparison between the CNMSSM and CMSSM.
\acknowledge
\bibliography{main}
\end{document}